\documentclass[aoas,nameyear,seceqn,dvips]{arximspdf}
\usepackage{dcolumn}
\usepackage{graphicx}

\doi{10.1214/10-AOAS422}
\volume{5}
\issue{2A}
\pubyear{2011}
\firstpage{798}
\lastpage{823}

\makeatletter
\newcolumntype{d}[1]{D{.}{.}{#1}}

\newcommand{\bmu}{\bolds{\mu}}
\newcommand{\btheta}{\bolds{\theta}}
\newcommand{\bbeta}{\bolds{\beta}}
\newcommand{\bepsilon}{\bolds{\epsilon}}
\newcommand{\bgamma}{\bolds{\gamma}}
\newcommand{\bSigma}{\bolds{\Sigma}}
\newcommand{\bPsi}{\bolds{\Psi}}
\newcommand{\bpi}{\bolds{\pi}}
\newcommand{\bOmega}{\bolds{\Omega}}
\newcommand{\bnu}{\bolds{\nu}}
\makeatother

\newproclaim{remark}{Remark}
\begin{document}
\begin{frontmatter}

\title{Mean--variance portfolio optimization when means and covariances
are unknown\thanksref{T1}}
\runtitle{Mean--variance portfolio optimization}
\thankstext{T1}{Supported by
NSF
Grants DMS-08-05879 at Stanford University and
DMS-09-06593 at SUNY, Stony Brook.}
\begin{aug}
\author[A]{\fnms{Tze Leung} \snm{Lai}\ead[label=e1]{lait@stanford.edu}},
\author[B]{\fnms{Haipeng} \snm{Xing}\corref{}\ead[label=e2]{xing@ams.sunysb.edu}}
\and
\author[C]{\fnms{Zehao} \snm{Chen}\ead[label=e3]{zehao.chen@gmail.com}}
\runauthor{T. L. Lai, H. Xing and Z. Chen}
\affiliation{Stanford University, SUNY at Stony Brook and Bosera Fund}
\address[A]{T. L. Lai\\
Department of Statistics\\
Stanford University\\
Palo Alto, California 94305\\
USA \\
\printead{e1}} 
\address[B]{T. Xing\\
Department of Applied Mathematics\\
\quad and Statistics\\
State University of New York\\
\quad at Stony Brook\\
Stony Brook, New York 11794\\
USA \\ \printead{e2}}
\address[C]{Z. Chen\\
Bosera Funds\\
Shenzhen\\
China \\
\printead{e3}}
\end{aug}

\received{\smonth{5} \syear{2009}}
\revised{\smonth{9} \syear{2010}}

%
\begin{abstract}
Markowitz's celebrated mean--variance portfolio optimization theory
assumes that the means and covariances of the underlying asset returns
are known. In practice, they are unknown and have to be estimated from
historical data. Plugging the estimates into the efficient frontier
that assumes known parameters has led to portfolios that may perform
poorly and have counter-intuitive asset allocation weights; this has been
referred to as the ``Markowitz optimization enigma.'' After reviewing
different approaches in the literature to address these difficulties, we
explain the root cause of the enigma and propose a new approach to resolve
it. Not only is the new approach shown to provide substantial
improvements over previous methods, but it also allows flexible modeling
to incorporate dynamic features and fundamental analysis of the training
sample of historical data, as illustrated in simulation and empirical studies.
\end{abstract}

%
\begin{keyword}
\kwd{Markowitz's portfolio theory}
\kwd{efficient frontier}
\kwd{empirical Bayes}
\kwd{stochastic optimization}.
\end{keyword}

\end{frontmatter}
%

\section{Introduction}\label{sec1}

The mean--variance (MV) portfolio optimization theory of Harry
Markowitz (\citeyear{Markowitz1952}, \citeyear{Markowitz1959}), Nobel laureate in economics, is widely
regarded as one of the foundational theories in financial economics.
It is a~single-period theory on the choice of portfolio weights
that provide the optimal tradeoff between the mean (as a measure of
profit) and the variance (as a measure of risk) of the portfolio return
for a future period. The theory, which will be briefly reviewed in the
next paragraph, assumes that the means and covariances of the underlying
asset returns are known. How to implement the theory in practice when
the means and covariances are unknown parameters has been an intriguing
statistical problem in financial economics. This paper proposes a novel
approach to resolve the long-standing problem and illustrates it with an
empirical study using CRSP (the Center for Research in Security Prices
of the University of Chicago) monthly stock price data, which can be
accessed via the Wharton Research Data Services at the University of
Pennsylvania.

For a portfolio consisting of $m$ assets (e.g., stocks)
with expected returns~$\mu_i$, let $w_i$ be the weight of
the portfolio's value invested in asset~$i$ such
that $\sum_{i=1}^m w_i=1$, and let
${\mathbf w}=(w_1, \dots, w_m)^T$, $\bmu=(\mu_1, \dots, \mu_m)^T$,
${\mathbf1}=(1, \dots, 1)^T$. The portfolio return has mean ${\mathbf w}^T
\bmu$
and variance ${\mathbf w}^T \bSigma{\mathbf w}$, whe\-re~$\bSigma$ is the covariance
matrix of the asset returns; see Lai and Xing (\citeyear{LX2008}), pages 67, 69--71.
Given a target value $\mu_*$ for the mean return
of a~portfolio, Markowitz characterizes an efficient portfolio by its weight
vector~${\mathbf w}_{\mathrm{eff}}$ that solves the optimization problem
\begin{equation}\label{eq11}
{\mathbf w}_{\mathrm{eff}} = \arg\min_{\mathbf w} {\mathbf w}^T \bSigma{\mathbf w}\quad
\mbox{subject to}\quad {\mathbf w}^T \bmu= \mu_*, {\mathbf w}^T {\mathbf1}=1,
{\mathbf w} \ge{\mathbf0}.
\end{equation}

 When short selling is allowed, the constraint ${\mathbf w} \ge{\mathbf0}$
(i.e., $w_i \ge0$ for all $i$) in~(\ref{eq11}) can be removed, yielding the
following problem that has an explicit solution:
\begin{eqnarray}\label{eq12}
{\mathbf w}_{\mathrm{eff}} &=&\arg\min_{{\mathbf w}: {\mathbf w}^T \bmu= \mu_*, \
{\mathbf w}^T {\mathbf1}=1 } {\mathbf w}^T \bSigma{\mathbf
w}\nonumber\\ [-8pt]\\ [-8pt]
&=& \{ B\bSigma^{-1} {\mathbf1} - A \bSigma^{-1} \bmu
+ \mu_* ( C\bSigma^{-1} \bmu- A \bSigma^{-1} {\mathbf1} )
\}/ D,\nonumber
\end{eqnarray}
where
$A= \bmu^T \bSigma^{-1} {\mathbf1} = {\mathbf1}^T \bSigma^{-1} \bmu,
B=\bmu^T \bSigma^{-1} \bmu, C= {\mathbf1}^T \bSigma^{-1} {\mathbf1}$, and
$D=\break BC-A^2.$

Markowitz's theory assumes known $\bmu$ and $\bSigma$. Since in practice
$\bmu$ and $\bSigma$ are unknown, a commonly used approach is to estimate
$\bmu$ and $\bSigma$ from historical data, under the assumption that
returns are i.i.d. A standard model for the price $P_{it}$ of the $i$th
asset at time $t$ in finance theory is geometric Brownian motion
$dP_{it}/P_{it}=\theta_i\,dt + \sigma_i\,dB^{(i)}_t$, where $\{ B^{(i)}_t,
t\ge0 \}$ is standard Brownian motion. The discrete-time analog
of this price process has returns $r_{it}=(P_{it}-P_{i,t-1}) / P_{i,t-1}$,
and log returns $\log(P_{it}/ P_{i,t-1})
= \log(1+r_{it}) \approx r_{it}$ that are i.i.d. $N(\theta_i - \sigma_i^2/2,
\sigma_i^2)$. Under the standard model, maximum likelihood estimates of
$\bmu$ and $\bSigma$ are the sample mean $\widehat{\bmu}$ and the sample
covariance matrix $\widehat{\bSigma}$, which are also method-of-moments
estimates without the assumption of normality and when the i.i.d.
assumption is replaced by weak stationarity (i.e., time-invariant
means and covariances).
It has been found, however, that replacing $\bmu$ and $\bSigma$ in
(\ref{eq11}) or (\ref{eq12}) by their sample counterparts $\widehat{\bmu}$ and
$\widehat{\bSigma}$ may perform poorly and a major direction in the
literature is to find other (e.g., Bayes and shrinkage) estimators
that yield better portfolios when they are plugged into (\ref{eq11}) or (\ref{eq12}). An
alternative method, introduced by Michaud (\citeyear{Michaud1989}) to tackle the ``Markowitz
optimization enigma,'' is to adjust the plug-in portfolio weights by
incorporating sampling variability of $(\widehat{\bmu}, \widehat
{\bSigma})$
via the bootstrap. Section \ref{sec2} gives a brief survey of these approaches.

Let ${\mathbf r}_t = (r_{1t}, \dots, r_{mt})^T$.
Since Markowitz's theory deals with portfolio returns in a future period,
it is more appropriate to use the conditional mean and covariance matrix
of the future returns ${\mathbf r}_{n+1}$ given the historical data~${\mathbf r}_n$,
${\mathbf r}_{n-1}, \dots$ based on a Bayesian model that forecasts the future
from the available data, rather than restricting to an i.i.d. model that
relates the future to the past via the unknown parameters $\bmu$ and
$\bSigma$ for future returns to be estimated from past data. More importantly,
this Bayesian formulation paves the way for a new approach that generalizes
Markowitz's portfolio theory to the case where the means and covariances
are unknown. When $\bmu$ and $\bSigma$ are estimated from data, their
uncertainties should be incorporated into the risk; moreover, it is not
possible to attain a target level of mean return as in Markowitz's
constraint ${\mathbf w}^T \bmu=\mu_*$ since $\bmu$ is unknown. To
address this
root cause of the Markowitz enigma, we introduce in Section \ref{sec3}
a~Bayesian approach that assumes a prior distribution for $(\bmu,
\bSigma)$
and formulates mean--variance portfolio optimization as a stochastic
optimization problem. This optimization problem reduces to that of
Markowitz when the prior distribution is degenerate. It
uses the posterior distribution
given current and past observations to incorporate the uncertainties of
$\bmu$ and $\bSigma$ into the variance of the portfolio return
${\mathbf w}^T {\mathbf r}_{n+1}$, where ${\mathbf w}$ is based on the posterior
distribution. The constraint in Markowitz's mean--variance
formulation can be included in the objective function by using a
Lagrange multiplier $\lambda^{-1}$ so that the optimization problem
is to evaluate the weight vector ${\mathbf w}$ that maximizes $E({\mathbf w}^T
{\mathbf r}_{n+1}) - \lambda\operatorname{Var} ({\mathbf w}^T {\mathbf r}_{n+1})$, for which
$\lambda$ can be regarded as a~risk aversion coefficient. To compare
with previous frequentist approaches that assume i.i.d. returns, Section
\ref{sec4} introduces a variant of the Bayes rule that uses bootstrap resampling
to estimate the performance criterion nonparametrically.

To apply this theory in practice, the investor has to figure out
his/her risk aversion coefficient, which may be a difficult task.
Markowitz's theory circumvents this by considering the \textit{efficient
frontier}, which is the $(\sigma, \mu)$ curve of efficient portfolios
as $\lambda$ varies over all possible values, where $\mu$ is the mean
and $\sigma^2$ the variance of the portfolio return. Investors, however,
often prefer to use $(\mu-\mu_0)/\sigma_e$, called the \textit{information
ratio}, as a measure of a portfolio's performance, where $\mu_0$ is
the expected return of a benchmark investment and $\sigma_e^2$ is the
variance of the portfolio's excess return over the benchmark portfolio;
see Grinold and Kahn (\citeyear{GK2000}), page 5. The benchmark
investment can be a market portfolio (e.g., S\&P500) or some other
reference portfolio, or a risk-free bank account with interest rate
$\mu_0$ (in which case the information ratio is often called the
\textit{Sharpe ratio}). Note that the information ratio is proportional to
$\mu-\mu_0$ and inversely proportional to $\sigma_e$, and can be
regarded as the excess return per unit of risk. In Section \ref{sec5} we
describe how $\lambda$ can be chosen for the rule developed in Section
\ref{sec3} to maximize the information ratio. Other statistical issues that
arise in practice are also considered in Sections~\ref{sec5} and~\ref{sec6} where they
lead to certain modifications of the basic rule. Among them are
dimension reduction when
$m$ (number of assets) is not small relative to $n$ (number of past periods
in the training sample) and departures of the historical data from the
working assumption of i.i.d. asset returns.
Section \ref{sec6} illustrates these methods in an empirical study in which
the rule thus obtained is compared with other rules proposed in the literature.
Some concluding remarks are given in Section \ref{sec7}.

\section{\texorpdfstring{Using better estimates of $\bmu$, $\bSigma$ or ${\mathbf w}_{\mathrm{eff}}$
to implement Markowitz's portfolio optimization theory}%
{Using better estimates of mu, Sigma or w eff
to implement Markowitz's portfolio optimization theory}}\label{sec2}

Since $\bmu$ and $\bSigma$ in Markowitz's efficient frontier are actually
unknown,
a natural idea is to replace them by the sample mean vector $\widehat
{\bmu}$
and covariance matrix $\widehat{\bSigma}$ of the training sample.
However, this plug-in frontier is no longer optimal because
$\widehat{\bmu}$ and $\widehat{\bSigma}$ actually differ from
$\bmu$ and~$\bSigma$, and Frankfurter, Phillips and Seagle (\citeyear{FPS1976})
and Jobson and Korkie (\citeyear{JK1980}) have reported that portfolios associated
with the plug-in frontier can
perform worse than an equally weighted portfolio that is highly
inefficient. Michaud (\citeyear{Michaud1989}) comments that the minimum variance
(MV) portfolio ${\mathbf w}_{\mathrm{eff}}$ based on $\widehat{\bmu}$ and
$\widehat{\bSigma}$ has serious deficiencies, calling the MV optimizers
``estimation-error maximizers.'' His argument is reinforced by
subsequent studies, for example, Best and Grauer (\citeyear{BG1991}), Chopra, Hensel and
Turner (\citeyear{CHT1993}), Canner et al. (\citeyear{CMW1997}), Simann (\citeyear{Simann1997})
 and Britten-Jones~(\citeyear{Britten1999}). Three approaches have emerged to address the difficulty
during the past two decades. The first approach uses multifactor models
to reduce the dimension in estimating $\bSigma$, and the second
approach uses Bayes or other shrinkage estimates of~$\bSigma$. Both
approaches use improved estimates of $\bSigma$ for the plug-in efficient
frontier. They have also been modified to provide better estimates of
$\bmu$, for example, in the quasi-Bayesian approach of Black and Litterman
(\citeyear{BL1990}). The third approach uses bootstrapping to correct
for the bias of $\widehat{\mathbf w}_{\mathrm{eff}}$ as an estimate of~${\mathbf w}_{\mathrm{eff}}$.

\subsection{Multifactor pricing models}\label{sec21}

Multifactor pricing models
relate the $m$ asset returns $r_i$ to $k$ factors $f_1, \dots, f_k$
in a regression model of the form
\begin{equation}\label{eq21}
r_i = \alpha_i + (f_1, \dots, f_k)^T \bbeta_i + \epsilon_i,
\end{equation}
in which $\alpha_i$ and $\bbeta_i$ are unknown regression
parameters and $\epsilon_i$ is an unobserved random disturbance
that has mean 0 and is uncorrelated with ${\mathbf f}:=(f_1, \dots, f_k)^T$.
The case $k=1$ is called a \textit{single-factor} (or \textit{single-index})
model. Under Sharpe's~(\citeyear{Sharpe1964}) capital asset pricing model (CAPM)
which assumes, besides known $\bmu$ and $\bSigma$, that the market
has a
risk-free asset with return $r_f$ (interest rate) and that all investors
minimize the variance of their portfolios for their target mean returns,
(\ref{eq21}) holds with $k=1$, $\alpha_i=r_f$
and $f=r_M-r_f$, where $r_M$ is the return of a hypothetical market
portfolio $M$ which can be approximated in practice by an index fund
such as Standard and Poor's (S\&P) 500 Index. The arbitrage pricing
theory (APT), introduced by Ross (\citeyear{Ross1976}), involves neither a market
portfolio nor a risk-free asset and states that a multifactor
model of the form (\ref{eq21}) should hold approximately in the absence of
arbitrage for sufficiently large $m$. The theory, however,
does not specify the factors and their number.
Methods for choosing factors in (\ref{eq21}) can be broadly classified
as economic and statistical, and commonly used statistical methods include
factor analysis and principal component analysis; see Section 3.4 of Lai
and Xing (\citeyear{LX2008}).

\subsection{Bayes and shrinkage estimators}\label{sec22}

A popular conjugate
family of prior distributions for estimation of covariance matrices
from i.i.d. normal random vectors~${\mathbf r}_t$ with mean $\bmu$ and
covariance matrix $\bSigma$ is
\begin{equation}\label{eq22}
\bmu|\bSigma\sim N(\bnu, \bSigma/ \kappa),\qquad
\bSigma\sim \mathit{IW}_m(\bPsi, n_0),
\end{equation}
where $\mathit{IW}_m(\bPsi, n_0)$ denotes the inverted Wishart
distribution with $n_0$ degrees of freedom and mean $\bPsi/(n_0-m-1)$.
The posterior distribution of $(\bmu, \bSigma)$
given $({\mathbf r}_1, \dots, {\mathbf r}_n)$ is also of the same form:
\[
\bmu| \bSigma\sim N\bigl(\widehat{\bmu}, \bSigma/(n+\kappa)\bigr),\qquad
\bSigma\sim \mathit{IW}_m \bigl( (n+n_0-m-1)\widehat{\bSigma}, n+n_0\bigr),
\]
where $\widehat{\bmu}$ and $\widehat{\bSigma}$ are
the Bayes estimators of $\bmu$ and $\bSigma$ given by
\begin{eqnarray}\label{eq23}
\widehat{\bmu} &=& \frac{\kappa}{n+ \kappa} \bnu+ \frac{n}{n+\kappa}
\bar{\mathbf r}, \nonumber\\
\widehat{\bSigma} &=&\frac{n_0-m-1}{n+n_0-m-1} \frac{\bPsi}{n_0-m-1}\nonumber\\ [-8pt]\\ [-8pt]
&&{}+ \frac{n}{n+n_0-m-1} \Biggl\{ \frac{1}{n} \sum_{i=1}^n
({\mathbf r}_t - \bar{\mathbf r})({\mathbf r}_t - \bar{\mathbf r})^T\nonumber\\
&&\hphantom{{}+ \frac{n}{n+n_0-m-1} \Biggl\{}{} + \frac{\kappa}{n+\kappa}
(\bar{\mathbf r} - \bnu)(\bar{\mathbf r} - \bnu)^T \Biggr\}.\nonumber
\end{eqnarray}
Note that the Bayes estimator $\widehat{\bSigma}$
adds to the MLE of $\bSigma$ the covariance matrix
$\kappa(\bar{\mathbf r} - \bnu)(\bar{\mathbf r} - \bnu)^T / (n+\kappa)$,
which accounts for the uncertainties due to replacing $\bmu$ by
$\bar{\mathbf r}$, besides shrinking this adjusted covariance matrix
toward the prior mean $\bPsi/ (n_0-m-1)$.

Simply using $\bar{\mathbf r}$ to estimate $\bmu$, Ledoit and Wolf (\citeyear{LW2003},
\citeyear{LW2004}) propose to shrink the MLE of $\bSigma$ toward a structured
covariance matrix, instead of using directly this Bayes estimator
which requires specification of the hyperparameters $\bmu$, $\kappa$,
$n_0$ and $\bPsi$. Their rationale is that whereas the MLE
${\mathbf S}= \sum_{t=1}^n ({\mathbf r}_t - \bar{\mathbf r})({\mathbf r}_t -
\bar{\mathbf r})^T / n$ has a large estimation error when $m(m+1)/2$
is comparable with $n$, a structured covariance matrix ${\mathbf F}$
has much fewer parameters that can be estimated with
smaller variances. They propose to estimate $\bSigma$
by a convex combination of $\widehat{\mathbf F}$ and ${\mathbf S}$:
\begin{equation}\label{eq24}
\widehat{\bSigma} = \hat{\delta} \widehat{\mathbf F} +
(1-\hat{\delta}) {\mathbf S},
\end{equation}
where $\hat{\delta}$ is an estimator of the \textit{optimal
shrinkage constant} $\delta$ used to shrink the MLE toward the
estimated structured covariance matrix $\widehat{\mathbf F}$.
Besides the covariance matrix~${\mathbf F}$ associated with a single-factor
model, they also suggest
using a~constant correlation model for ${\mathbf F}$ in which all
pairwise correlations are identical, and have found that it
gives comparable performance in simulation and empirical studies. They
advocate using this shrinkage estimate in lieu of
${\mathbf S}$ in implementing Markowitz's efficient frontier.

The difficulty of estimating $\bmu$ well enough for the plug-in
portfolio to have reliable performance was pointed out by Black and
Litterman (\citeyear{BL1990}), who proposed the following pragmatic quasi-Bayesian
approach to address this difficulty. Whereas Jorion (\citeyear{Jorion1986}) had used
earlier a shrinkage estimator similar to $\widehat{\bmu}$ in (\ref{eq23}),
which can be viewed as shrinking a prior mean~$\bnu$ to the sample
mean $\bar{\mathbf r}$ (instead of the other way around), Black and
Litterman's approach basically amounted to shrinking an investor's
subjective estimate of $\bmu$ to the market's estimate implied by an
``equilibrium portfolio.'' The investor's subjective guess of $\bmu$
is described in terms of ``views'' on linear combinations of asset
returns, which can be based on past observations and the investor's
personal/expert opinions. These views are represented by ${\mathbf P}\bmu
\sim N({\mathbf q}, \bOmega)$, where ${\mathbf P}$ is a $p\times m$ matrix of the
investor's ``picks'' of the assets to express the guesses, and
$\bOmega$ is a diagonal matrix that expresses the investor's
uncertainties in the views via their variances. The ``equilibrium
portfolio,'' denoted by $\widetilde{\mathbf w}$, is based on a normative
theory of
an equilibrium market, in which $\widetilde{\mathbf w}$ is assumed to
solve the mean--variance optimization problem $\max_{\mathbf w} (
{\mathbf w}^T \bpi- \lambda{\mathbf w}^T \bSigma{\mathbf w} )$, with
$\lambda$ being the average risk-aversion level of the market and
$\bpi$ representing the market's view of $\bmu$. This theory yields the
relation $\bpi= 2 \lambda\bSigma\widetilde{\mathbf w}$, which can be used
to infer $\bpi$ from the market capitalization or benchmark portfolio as
a surrogate of $\widetilde{\mathbf w}$. Incorporating uncertainty in the
market's view of $\bmu$, Black and Litterman assume that $\bpi- \bmu
\sim N({\mathbf0}, \tau\bSigma)$, in which $\tau\in(0,1)$ is a~small
parameter, and also set exogenously $\lambda=1.2$; see Meucci~(\citeyear{Meucci2010}).
 Combining ${\mathbf P}\bmu\sim N({\mathbf q}, \bOmega)$ with
$\bpi-\bmu\sim N({\mathbf0}, \tau\bSigma)$ under a working independence
assumption between the two multivariate normal distributions yields the
Black--Litterman estimate of $\bmu$:
\begin{equation}\label{eq25}
\widehat{\bmu}^{\mathrm{BL}} = [ (\tau\bSigma)^{-1} + {\mathbf P}^T
\bOmega^{-1} {\mathbf P} ]^{-1} [ (\tau\bSigma)^{-1} \bpi+
{\mathbf P}^T \bOmega^{-1} {\mathbf q} ],
\end{equation}
with covariance matrix $[ (\tau\bSigma)^{-1} + {\mathbf P}^T
\bOmega^{-1} {\mathbf P} ]^{-1}$. Various modifications and extensions
of their idea have been proposed; see Meucci (\citeyear{Meucci2005}), pages 426--437,
Fabozzi et al. (\citeyear{FKPF2007}), pages 232--253, and Meucci (\citeyear{Meucci2010}). These extensions
have the basic form (\ref{eq25}) or some variant thereof, and differ mainly
in the normative model used to generate an equilibrium portfolio. Note
that~(\ref{eq25}) involves $\bSigma$, which Black and Litterman estimated by
using the sample covariance matrix of historical data, and that
their focus was to address the
estimation of $\bmu$ for the plug-in portfolio. Clearly Bayes or
shrinkage estimates of $\bSigma$ can be used instead.

\subsection{Bootstrapping and the resampled frontier}\label{sec23}

To adjust for the bias of~$\widehat{\mathbf w}_{\mathrm{eff}}$ as an estimate
of ${\mathbf w}_{\mathrm{eff}}$,
Michaud (\citeyear{Michaud1989}) uses the average of the bootstrap weight vectors:
\begin{equation}\label{eq26}
\bar{\mathbf w} = B^{-1} \sum_{b=1}^B \widehat{\mathbf w}_b^*,
\end{equation}
where $\widehat{\mathbf w}_b^*$ is the estimated optimal portfolio weight
vector based on the $b$th bootstrap sample $\{ {\mathbf r}_{b1}^*, \dots,
{\mathbf r}_{bn}^* \}$ drawn with replacement from the observed sample
$\{ {\mathbf r}_1, \dots, {\mathbf r}_n \}$. Specifically,
the $b$th bootstrap sample has sample mean vector $\widehat{\bmu}_b^*$
and covariance matrix $\widehat{\bSigma}_b^*$, which can be used to replace
$\bmu$ and $\bSigma$ in (\ref{eq11}) or (\ref{eq12}), thereby yielding
$\widehat{\mathbf w}_b^*$. Thus, the resampled efficient frontier
corresponds to plotting $\bar{\mathbf w}^T \widehat{\bmu}$ versus
$\sqrt{\bar{\mathbf w}^T \widehat{\bSigma} \bar{\mathbf w}}$ for a fine grid
of $\mu_*$ values, where $\bar{\mathbf w}$ is defined by (\ref{eq26}) in which
$\widehat{\mathbf w}_b^*$ depends on the target level $\bmu_*$.

\section{A stochastic optimization approach}\label{sec3}

The Bayesian and shrinkage methods in Section \ref{sec22} focus primarily on
Bayes estimates of $\bmu$ and $\bSigma$ (with normal and inverted
Wishart priors) and shrinkage estimators of $\bSigma$. However,
the construction of efficient portfolios when $\bmu$ and $\bSigma$
are unknown is more complicated than trying to estimate them as well as
possible and then plugging the estimates into (\ref{eq11}) or (\ref{eq12}).
Note in this connection that (\ref{eq12}) involves $\bSigma^{-1}$
instead of $\bSigma$ and that estimating $\bSigma$ as well as possible
does not imply that $\bSigma^{-1}$ is reliably estimated.
Estimation of a high-dimensional $m\times m$ covariance matrix and its
inverse when $m^2$ is not small compared to $n$ has been recognized as
a difficult statistical problem and attracted much recent attention; see,
for example, Ledoit and Wolf (\citeyear{LW2004}), Huang et al. (\citeyear{HLPL2006}), Bickel and
Lavina (\citeyear{BL2008}) and Fan, Fan and Lv (\citeyear{FFL2008}). Some sparsity condition or a
low-dimensional factor structure is
needed to obtain an estimate which is close to $\bSigma$ and whose
inverse is close to $\bSigma^{-1}$, but the conjugate prior family
(\ref{eq22}) that motivates the (linear) shrinkage estimators~(\ref{eq23})
or (\ref{eq24}) does not reflect such sparsity. For high-dimensional
weight vectors $\widehat{\mathbf w}_{\mathrm{eff}}$, direct application of the
bootstrap for bias correction is also problematic.

A major difficulty with the ``plug-in'' efficient frontier (which uses
${\mathbf S}$ to estimate $\bSigma$ and $\bar{\mathbf r}$ to estimate $\bmu$),
its variants that estimate $\bSigma$ by (\ref{eq24}) and~$\bmu$ by (\ref{eq23}) or
the Black--Litterman method, and its
``resampled'' version is that Markowitz's idea of using the variance of
${\mathbf w}^T {\mathbf r}_{n+1}$ as a measure of the portfolio's risk cannot be
captured simply by the plug-in estimates ${\mathbf w}^T \widehat{\bSigma}
{\mathbf w}$ of $\operatorname{Var}({\mathbf w}^T {\mathbf r}_{n+1})$ and ${\mathbf w}^T \widehat
{\bmu}$ of
$E({\mathbf w}^T {\mathbf r}_{n+1})$. This difficulty was recognized by
Broadie (\citeyear{Broadie1993}), who used the terms \textit{true frontier} and
\textit{estimated frontier} to refer to Markowitz's efficient frontier (with
known $\bmu$ and $\bSigma$) and the plug-in efficient frontier,
respectively, and who also suggested considering the \textit{actual} mean
and variance of the return of an estimated frontier portfolio.
Whereas the problem of minimizing $\operatorname{Var}({\mathbf w}^T {\mathbf r}_{n+1})$
subject to a given level~$\mu_*$ of the mean return $E({\mathbf w}^T {\mathbf
r}_{n+1})$ is meaningful in Markowitz's framework,
in which both $E({\mathbf r}_{n+1})$ and $\operatorname{Cov}({\mathbf r}_{n+1})$ are known,
the surrogate problem of minimizing ${\mathbf w}^T \widehat{\bSigma} {\mathbf w}$
under the constraint ${\mathbf w}^T \widehat{\bmu} =\mu_*$
ignores the fact that both $\widehat{\bmu}$ and $\widehat{\bSigma}$
have inherent errors (risks) themselves. In this section we
consider the more fundamental problem
\begin{equation}\label{eq31}
\max \{ E({\mathbf w}^T {\mathbf r}_{n+1}) - \lambda\operatorname{Var}({\mathbf w}^T
{\mathbf r}_{n+1}) \}
\end{equation}
when $\bmu$ and
$\bSigma$ are unknown and treated as state variables whose
uncertainties are specified by their posterior distributions given
the observations ${\mathbf r}_1, \ldots, {\mathbf r}_n$ in a~Bayesian framework.
The weights ${\mathbf w}$ in (\ref{eq31}) are random vectors that depend on ${\mathbf r}_1,\ldots,{\mathbf r}_n$.
Note that if the prior distribution puts all its mass at $(\bmu_0,
\bSigma_0)$,
then the minimization problem (\ref{eq31}) reduces to Markowitz's portfolio
optimization problem that assumes $\bmu_0$ and $\bSigma_0$ are
given. The Lagrange multiplier $\lambda$ in (\ref{eq31}) can be regarded as
the investor's risk-aversion index when variance is used to measure
risk.

\subsection{\texorpdfstring{Solution of the optimization problem (\protect\ref{eq31})}%
{Solution of the optimization problem (3.1)}}\label{sec31}

The problem (\ref{eq31}) is not a~standard stochastic optimization problem
because of the term $[ E({\mathbf w}^T
{\mathbf r}_{n+1}) ]^2$ in $\operatorname{Var}({\mathbf w}^T {\mathbf r}_{n+1}) =
E[({\mathbf w}^T {\mathbf r}_{n+1})^2 ] - [ E({\mathbf w}^T {\mathbf r}_{n+1})
]^2$. A standard stochastic optimization problem in the Bayesian
setting is of the form $\max_{a \in{\mathcal A}} E g({\mathbf X}, \btheta,
a)$, in which $g({\mathbf X}, \btheta, a)$ is the reward when action $a$
is taken, ${\mathbf X}$ is a random vector with distribution
$F_{\btheta}$, $\btheta$ has a prior distribution and the maximization
is over the action space ${\mathcal A}$. The key to its solution is the
law of conditional expectations $Eg({\mathbf X}, \btheta, a) = E \{
E[ g({\mathbf X}, \btheta, a) | {\mathbf X} ]\}$, which
implies that the stochastic optimization problem can be solved by
choosing $a$ to maximize the posterior reward $E[ g({\mathbf X},
\btheta, a) | {\mathbf X} ] \}$. This key idea, however,
cannot be applied to the problem of maximizing or minimizing nonlinear
functions of $Eg({\mathbf X}, \btheta, a)$, such as $[ Eg({\mathbf X},
\btheta, a)]^2$ that is involved in~(\ref{eq31}).

Our method of solving (\ref{eq31}) is to
convert it to a standard stochastic control problem
by using an additional parameter. Let $W={\mathbf w}^T{\mathbf r}_{n+1}$ and
note that $E(W)-\lambda\operatorname{Var}(W) =h(EW, EW^2)$, where
$h(x, y)=x+\lambda x^2-\lambda y$. Let $W_B= {\mathbf w}_B^T {\mathbf r}_{n+1}$
and $\eta= 1+2\lambda E(W_B)$, where ${\mathbf w}_B$ is the Bayes weight
vector. Then
\begin{eqnarray*}
0 &\ge& h(EW, EW^2) - h(EW_B, EW_B^2) \\
& = & E(W)-E(W_B)-\lambda\{ E(W^2)-E(W_B^2) \} + \lambda\{ (EW)^2
- (EW_B)^2 \} \\
&=& \eta\{ E(W)-E(W_B)\} + \lambda\{ E(W_B^2)- E(W^2) \} + \lambda
\{ E(W)-E(W_B) \}^2 \\
& \ge &\{ \lambda E(W_B^2)-\eta E(W_B) \} - \{ \lambda E(W^2) - \eta
E(W) \}.
\end{eqnarray*}
Moreover, the last inequality is strict unless $E W = EW_B$, in which
case the first inequality is strict unless $EW^2
= EW^2_B$. This shows that the last term above is ${\le}0$ or,
equivalently,
\begin{equation}\label{eq32}
\lambda E(W^2) - \eta E(W) \ge\lambda E(W_B^2) - \eta E(W_B),
\end{equation}
and that equality holds in (\ref{eq32}) if and only if $W$ has the same
mean and variance as
$W_B$. Hence, the stochastic optimization problem (\ref{eq31}) is equivalent to
minimizing $\lambda E [ ({\mathbf w}^T {\mathbf r}_{n+1})^2] - \eta E({\mathbf w}^T
{\mathbf r}_{n+1})$ over weight vectors ${\mathbf w}$ that can depend on ${\mathbf r}_1,
\ldots, {\mathbf r}_n$. Since $\eta=1+2 \lambda E(W_B)$ is a linear
function of the
solution of (\ref{eq31}), we cannot apply this equivalence directly to the
unknown~$\eta$. Instead we solve a~family of standard stochastic optimization
problems over $\eta$ and then choose the $\eta$ that maximizes the reward
in (\ref{eq31}).

To summarize, we can solve (\ref{eq31}) by rewriting it as the following
maximization problem over $\eta$:
\begin{equation}\label{eq33}
\max_{\eta} \{ E[ {\mathbf w}^T(\eta) {\mathbf r}_{n+1} ] - \lambda
\operatorname{Var}
[ {\mathbf w}^T (\eta) {\mathbf r}_{n+1}] \},
\end{equation}
where ${\mathbf w}(\eta)$ is the solution of the stochastic
optimization problem
\[
{\mathbf w}(\eta) = \arg\min_{{\mathbf w}} \{ \lambda E[ ({\mathbf w}^T
{\mathbf r}_{n+1})^2]- \eta E({\mathbf w}^T {\mathbf r}_{n+1}) \}.
\]

\subsection{Computation of the optimal weight vector}\label{sec32}

$\!\!$Let
$\bmu_n$ and ${\mathbf V}_n$ be the~pos\-terior mean and second moment matrix
given the set ${\mathcal R}_n$ of current and past returns ${\mathbf r}_1,
\dots,
{\mathbf r}_n$. Since ${\mathbf w}$ is based on ${\mathcal R}_n$, it follows from
$E({\mathbf r}_{n+1}|{\mathcal R}_n) = \bmu_n$ and $E({\mathbf r}_{n+1}{\mathbf r}_{n+1}^T
|{\mathcal R}_n) = {\mathbf V}_n$ that
\begin{equation}\label{eq34}
E({\mathbf w}^T{\mathbf r}_{n+1}) = E( {\mathbf w}^T \bmu_n ),\qquad
E[({\mathbf w}^T{\mathbf r}_{n+1})^2] = E( {\mathbf w}^T {\mathbf V}_n {\mathbf w}).
\end{equation}
Without short selling, the weight vector ${\mathbf w}(\eta)$ in (\ref{eq33})
is given by the following analog of (\ref{eq11}):
\begin{equation}\label{eq35}
{\mathbf w}(\eta) = \arg\min_{{\mathbf w}: {\mathbf w}^T {\mathbf1}=1,
{\mathbf w}\ge{\mathbf0}} \{ \lambda{\mathbf w}^T {\mathbf V}_n {\mathbf w} - \eta
{\mathbf w}^T \bmu_n \},
\end{equation}
which can be computed by quadratic programming (e.g., by
\texttt{quadprog} in\break \texttt{MATLAB}). When short selling is allowed
but there are limits on short setting, the constraint ${\mathbf w}\ge{\mathbf
0}$ can be replaced by ${\mathbf w} \ge{\mathbf w}_0$, where ${\mathbf w}_0$ is
a vector of negative numbers. When there is no limit on short selling,
the constraint ${\mathbf w}\ge{\mathbf0}$ in (\ref{eq35}) can be removed and ${\mathbf
w}(\eta)$ in (\ref{eq33}) is given explicitly by
\begin{eqnarray}\label{eq36}
{\mathbf w}(\eta) &=& \arg\min_{{\mathbf w}: {\mathbf w}^T {\mathbf1}=1}
\{ \lambda{\mathbf w}^T {\mathbf V}_n {\mathbf w} - \eta{\mathbf w}^T \bmu_n
\}\nonumber\\ [-8pt]\\ [-8pt]
&=& \frac{1}{C_n} {\mathbf V}_n^{-1} {\mathbf1} +
\frac{\eta}{2\lambda} {\mathbf V}_n^{-1} \biggl( \bmu_n - \frac{A_n
}{C_n} {\mathbf1} \biggr),\nonumber
\end{eqnarray}
where the second equality can be derived by using a Lagrange
multiplier~and
\begin{equation}\label{eq37}
\hspace*{20pt}A_n=\bmu_n^T {\mathbf V}_n^{-1} {\mathbf1}={\mathbf1}^T {\mathbf V}_n^{-1}\bmu_n,\qquad
B_n=\bmu_n^T {\mathbf V}_n^{-1} \bmu_n,\qquad
C_n = {\mathbf1}^T {\mathbf V}_n^{-1} {\mathbf1}.
\end{equation}
 Quadratic programming can be used to compute ${\mathbf w}(\eta)$ for more
general linear and quadratic constraints than those in (\ref{eq35}); see
Fabozzi et al. (\citeyear{FKPF2007}), pages 88--92.

Note that (\ref{eq35}) or (\ref{eq36}) essentially plugs the Bayes estimates of
$\bmu$ and ${\mathbf V} := \bSigma+ \bmu\bmu^T$ into the optimal weight
vector that assumes $\bmu$ and $\bSigma$ to be known. However, unlike
the ``plug-in'' efficient
frontier described in the first paragraph of Section \ref{sec2}, we have first
transformed the original mean--variance portfolio optimization problem
into a ``mean versus second moment'' optimization problem that has an
additional parameter $\eta$. Putting (\ref{eq35}) or (\ref{eq36}) into
\begin{equation}\label{eq38}
C(\eta):= E[ {\mathbf w}^T(\eta) \bmu_n ] + \lambda( E
[ {\mathbf w}^T(\eta) \bmu_n ] )^2 - \lambda E [ {\mathbf w}^T(\eta)
{\mathbf V}_n {\mathbf w}(\eta) ],
\end{equation}
 which is equal to $E[{\mathbf w}^T(\eta) {\mathbf r} ] - \lambda
\operatorname{Var}[{\mathbf w}^T(\eta) {\mathbf r}]$ by (\ref{eq34}), we can use Brent's
method [Press et al. (\citeyear{PTWF1992}), pages 359--362] to maximize $C(\eta)$. It should
be noted
that this argument implicitly assumes that the maximum of (\ref{eq31}) is
attained by some ${\mathbf w}$ and is finite. Whereas this assumption
is satisfied when there are limits on short selling as in (\ref{eq35}), it
may not hold when there is no limit on short selling. In fact, the explicit
formula of ${\mathbf w}(\eta)$ in (\ref{eq36}) can be used to express (\ref{eq38}) as
a quadratic function of $\eta$:
\begin{eqnarray*}
C(\eta)& =& \frac{\eta^2}{4 \lambda} E \biggl\{\!\! \biggl(B_n-\frac
{A_n^2}{C_n} \biggr)\!\biggl(B_n-\frac{A_n^2}{C_n}-1 \biggr)\!\! \biggr\} + \eta E
\biggl\{\!\!
\biggl(\frac{1}{2 \lambda}+\frac{A_n}{C_n} \biggr)\!\!\biggl(B_n-\frac{A_n^2}{C_n}
\biggr)\!\! \biggr\} \\
&&{}+ E \biggl\{ \frac{A_n}{C_n} + \lambda\frac{A_n^2-C_n}{C_n^2}\biggr\},
\end{eqnarray*}
 which has a maximum only if
\begin{equation}\label{eq39}
E\biggl\{\!\! \biggl(B_n-\frac{A_n^2}{C_n} \biggr)
\biggl(B_n-\frac{A_n^2}{C_n}-1 \biggr)\!\! \biggr\} <0.
\end{equation}
 In the case $E\{ (B_n-\frac{A_n^2}{C_n} ) (B_n
-\frac{A_n^2}{C_n}-1 ) \} >0$, $C(\eta)$ has\vspace*{1pt} a minimum
instead and
approaches to $\infty$ as $|\eta| \rightarrow\infty$. In this case,
(\ref{eq31}) has an infinite value and should be defined as a supremum (which
is not attained) instead of a~maximum.

\begin{remark*}
Let $\bSigma_n$ denote the posterior covariance matrix
given ${\mathcal R}_n$. Note that the law of iterated
conditional expectations, from which (\ref{eq34}) follows, has the following
analog for $\operatorname{Var}(W)$:
\begin{eqnarray}\label{eq310}
\operatorname{Var}(W) &=& E[ \operatorname{Var}(W|{\mathcal R}_n ) ] +
\operatorname{Var}[ E(W|{\mathcal R}_n ) ]\nonumber\\ [-8pt]\\ [-8pt]
&=& E({\mathbf w}^T \bSigma_n {\mathbf w}) + \operatorname{Var}({\mathbf w}^T \bmu_n).\nonumber
\end{eqnarray}
Using $\bSigma_n$ to replace $\bSigma$ in the optimal weight vector
that assumes $\bmu$ and~$\bSigma$ to be known, therefore, ignores
the variance of ${\mathbf w}^T\bmu_n$ in (\ref{eq310}), and this
omission is an important root cause for the Markowitz optimization
enigma related to ``plug-in'' efficient frontiers.
\end{remark*}

\section{Empirical Bayes, bootstrap approximation and frequentist
risk}\label{sec4}

For more flexible modeling, one can allow the prior distribution in the
preceding Bayesian approach to include unspecified
hyperparameters, which can be estimated from the training sample by
maximum likelihood, or method of moments or other methods. For example,
for the conjugate prior (\ref{eq22}), we can assume $\bnu$ and $\bPsi$ to
be functions of certain hyperparameters that are associated with
a multifactor model of the type (\ref{eq21}). This amounts to using an empirical
Bayes model for $(\bmu, \bSigma)$ in the stochastic optimization
problem~(\ref{eq31}). Besides a prior distribution for $(\bmu, \bSigma)$,
(\ref{eq31}) also requires specification of the common distribution of the
i.i.d. returns to evaluate $E_{\bmu, \bSigma}({\mathbf w}^T {\mathbf r}_{n+1})$
and $\operatorname{Var}_{\bmu, \bSigma}({\mathbf w}^T {\mathbf r}_{n+1})$. The bootstrap provides
a~nonparametric method to evaluate these quantities, as described below.

\subsection{Bootstrap estimate of performance}\label{sec41}

To begin with, note that we can evaluate the frequentist performance
of asset allocation rules by making use of the bootstrap method. The
bootstrap samples $\{ {\mathbf r}_{b1}^*, \ldots, {\mathbf r}_{bn}^* \}$ drawn
with replacement from the observed sample $\{ {\mathbf r}_1, \ldots, {\mathbf r}_n \}$,
$1\le b \le B$, can be used to estimate its\vspace*{1pt} $E_{\bmu, \bSigma}( {\mathbf w}_n^T
{\mathbf r}_{n+1})= E_{\bmu, \bSigma}({\mathbf w}_n^T \bmu)$ and $\operatorname{Var}_{\bmu,
\bSigma}({\mathbf w}_n^T {\mathbf r}_{n+1}) = E_{\bmu, \bSigma}({\mathbf w}_n^T
\bSigma{\mathbf w}_n)+\operatorname{Var}_{\bmu, \bSigma}({\mathbf w}_n^T \bmu)$
of various portfolios $\Pi$ whose weight vectors~${\mathbf w}_n$ may
depend on ${\mathbf r}_1, \dots, {\mathbf r}_n$.
In particular, we can use Bayes or other
estimators for $\bmu_n$ and ${\mathbf V}_n$ in (\ref{eq35}) or (\ref{eq36}) and
then choose $\eta$ to maximize the bootstrap estimate of
$E_{\bmu, \bSigma}( {\mathbf w}_n^T {\mathbf r}_{n+1}) - \lambda\operatorname{Var}_{\bmu,
\bSigma}({\mathbf w}_n^T {\mathbf r}_{n+1})$. This is tantamount to using the
empirical distribution of ${\mathbf r}_1, \dots, {\mathbf r}_n$ to be the
common distribution of the returns. In particular, using $\bar{\mathbf r}$
for $\bmu_n$ in (\ref{eq35}) and the second moment matrix $n^{-1}\sum_{t=1}^n
{\mathbf r}_t {\mathbf r}_t^T$ for ${\mathbf V}_n$ in~(\ref{eq36}) provides a
``nonparametric empirical Bayes'' variant, abbreviated by NPEB
hereafter, of the optimal rule in Section \ref{sec3}.

\subsection{A simulation study of Bayes and frequentist
rewards}\label{sec42}

The following simulation study assumes i.i.d. annual returns (in \%) of
$m=4$ assets whose mean vector and covariance matrix are generated from
the normal and inverted Wishart prior distribution (\ref{sec22}) with $\kappa=5$,
$n_0 = 10$, $\bnu=(2.48, 2.17, 1.61, 3.42)^T$
and the hyperparameter $\bPsi$ given by
\begin{eqnarray*}
\bPsi_{11}&=&3.37,\qquad \bPsi_{22}=4.22,\qquad \bPsi_{33}=2.75,\qquad
\bPsi_{44}=8.43,\\
\bPsi_{12}&=&2.04, \\
\bPsi_{13}&=&0.32,\qquad \bPsi_{14}=1.59,\qquad \bPsi_{23}=-0.05,\\
\bPsi_{24}&=&3.02,\qquad \bPsi_{34}=1.08.
\end{eqnarray*}
 We consider four scenarios for the case $n=6$ without short selling.
The first scenario assumes this prior distribution and studies the Bayesian
reward for $\lambda=1, 5$ and 10. The other scenarios consider the
frequentist reward at three values of $(\bmu, \bSigma)$ generated from
the prior distribution. These values, denoted by Freq 1, Freq 2, Freq 3,
are as follows:
\begin{longlist}
\item[Freq 1:] $\bmu=(2.42, 1.88, 1.58, 3.47)^T$, $\bSigma_{11}=1.17,
\bSigma_{22}=0.82, \bSigma_{33}=1.37,\break \bSigma_{44}=2.86,
\bSigma_{12}=0.79, \bSigma_{13}=0.84, \bSigma_{14}=1.61,
\bSigma_{23}=0.61, \bSigma_{24}=1.23, \bSigma_{34}=1.35$.

\item[Freq 2:] $\bmu=(2.59, 2.29, 1.25, 3.13)^T$, $\bSigma_{11}=1.32,
\bSigma_{22}=0.67, \bSigma_{33}=1.43,\break \bSigma_{44}=1.03,
\bSigma_{12}=0.75, \bSigma_{13}=0.85, \bSigma_{14}=0.68,
\bSigma_{23}=0.32, \bSigma_{24}=0.44, \bSigma_{34}=0.61$.

\item[Freq 3:] $\bmu=(1.91, 1.58, 1.03, 2.76)^T$, $\bSigma_{11}=1.00,
\bSigma_{22}=0.83, \bSigma_{33}=0.35,\break \bSigma_{44}=0.62,
\bSigma_{12}=0.73, \bSigma_{13}=0.26, \bSigma_{14}=0.36,
\bSigma_{23}=0.16, \bSigma_{24}=0.50, \bSigma_{34}=0.14$.
\end{longlist}

%
\begin{table}
\caption{Rewards of four portfolios formed from $m=4$ assets}\label{table:4assets}
\tabcolsep=-2pt
\begin{tabular*}{\textwidth}{@{\extracolsep{\fill}}lccccc@{}}
\hline
$\bolds\lambda$ & $\bolds{(}\bmu\bolds{,} \bSigma\bolds{)}$ & \textbf{Bayes} &\textbf{Plug-in}
&\textbf{Oracle} & \textbf{NPEB} \\
\hline
\phantom{0}1 & Bayes & 0.0324 (2.47e$-$5) & 0.0317 (2.55e$-$5) & 0.0328 (2.27e$-$5)
& 0.0324 (2.01e$-$5) \\
& Freq 1 & 0.0332 (2.61e$-$6) & 0.0324 (5.62e$-$6) & 0.0332\phantom{ (2.27e$-$5)} & 0.0332
(2.56e$-$6) \\
& Freq 2 & 0.0293 (7.23e$-$6) & 0.0282 (5.32e$-$6) & 0.0298\phantom{ (2.27e$-$5)} & 0.0293
(7.12e$-$6) \\
& Freq 3 & 0.0267 (4.54e$-$6) & 0.0257 (5.57e$-$6) & 0.0268\phantom{ (2.27e$-$5)} & 0.0267
(4.73e$-$6) \\ [3pt]
\phantom{0}5 & Bayes & 0.0262 (2.33e$-$5) & 0.0189 (1.21e$-$5) & 0.0267 (2.02e$-$5)
& 0.0262 (1.89e$-$5) \\
&Freq 1& 0.0272 (4.06e$-$6) & 0.0182 (5.54e$-$6) & 0.0273\phantom{ (2.02e$-$5)} & 0.0272
(2.60e$-$6) \\
&Freq 2& 0.0233 (9.35e$-$6) & 0.0183 (3.88e$-$6) & 0.0240\phantom{ (2.02e$-$5)} & 0.0234
(1.03e$-$5)\\
&Freq 3& 0.0235 (5.25e$-$6) & 0.0159 (2.88e$-$6) & 0.0237\phantom{ (2.02e$-$5)} & 0.0235 (5.27e$-$6)
\\ [3pt]
10 & Bayes & 0.0184 (2.54e$-$5) & 0.0067 (7.16e$-$6) & 0.0190 (2.08e$-$5)
& 0.0183 (2.23e$-$5) \\
&Freq 1& 0.0197 (7.95e$-$6) & 0.0063 (3.63e$-$6) & 0.0199\phantom{ (2.08e$-$5)} & 0.0198
(4.19e$-$6)\\
&Freq 2& 0.0157 (1.08e$-$5) & 0.0072 (3.00e$-$6) & 0.0168\phantom{ (2.08e$-$5)} & 0.0159
(1.13e$-$5) \\
&Freq 3& 0.0195 (6.59e$-$6) & 0.0083 (1.62e$-$6) & 0.0198\phantom{ (2.08e$-$5)} & 0.0196 (5.95e$-$6)
\\ \hline
\end{tabular*}
\end{table}

%
\begin{figure}

\includegraphics{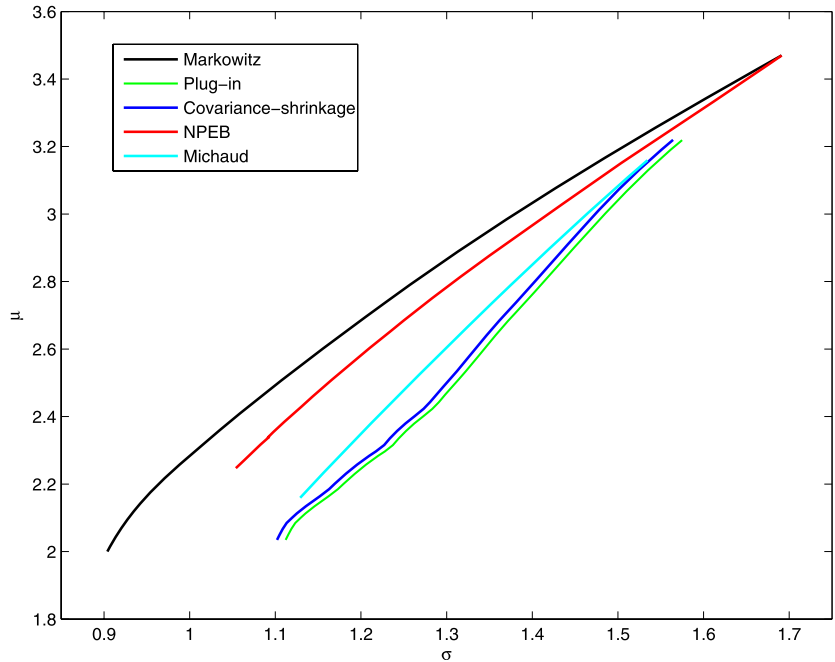}

\caption{$(\sigma, \mu)$ curves of different portfolios.}\label{effi}
\end{figure}

Table \ref{table:4assets} compares the Bayes rule that maximizes
(\ref{eq31}), called ``Bayes'' hereafter, with three other rules: (a) the
``oracle'' rule that assumes $\bmu$ and~$\bSigma$ to be known, (b)~the
plug-in rule that replaces $\bmu$ and $\bSigma$ by the sample
estimates of $\bmu$ and $\bSigma$, and (c) the NPEB (nonparametric empirical
Bayes) rule described in Section \ref{sec41} Note that although
both (b) and (c) use the sample mean vector and sample covariance (or
second moment) matrix, (b)~simply plugs the sample estimates into the
oracle rule while (c) uses the empirical distribution to replace the
common distribution of the returns in the Bayes rule.
For the plug-in rule, the quadratic
programming procedure may have numerical difficulties if the sample
covariance matrix is nearly singular. If it should happen, we
use the default option of adding 0.005${\mathbf I}$ to the
sample covariance matrix. Each result in Table \ref{table:4assets}
is based on 500
simulations, and the standard errors are given in parentheses.
In each scenario, the reward of the NPEB rule is close to that of
the Bayes rule and somewhat smaller than that of the oracle rule. The
plug-in rule has substantially smaller rewards, especially for larger
values of $\lambda$.

\subsection{\texorpdfstring{Comparison of the $(\sigma, \mu)$ plots of different portfolios}%
{Comparison of the (sigma, mu) plots of different portfolios}}\label{sec43}

The set of points in the $(\sigma, \mu)$ plane that correspond to the
returns of portfolios of the $m$ assets is called the \textit{feasible
region}. As $\lambda$ varies over $(0, \infty)$, the $(\sigma, \mu)$
values of the oracle rule correspond to Markowitz's efficient frontier
which assumes known $\bmu$ and $\bSigma$ and which is the upper left
boundary of the feasible region. For portfolios whose weights do not
assume knowledge of $\bmu$ and $\bSigma$, the $(\sigma, \mu)$
values lie
on the right of Markowitz's efficient frontier.
Figure~\ref{effi} plots the $(\sigma, \mu)$ values of different portfolios
formed from $m=4$ assets without short selling and a training sample of
size $n=6$ when $(\bmu, \bSigma)$ is given by the frequentist
scenario Freq 1 above. Markowitz's efficient frontier is computed
analytically by varying $\mu_*$ in (\ref{eq11}) over a grid of values.
The $(\sigma, \mu)$ curves of the plug-in, covariance-shrinkage
[Ledoit and Wolf~(\citeyear{LW2004})] and
Michaud's resampled portfolios are computed by Monte Carlo,
using 500 simulated paths, for each value of $\mu_*$ in a grid
ranging from 2.0 to 3.47. The $(\sigma, \mu)$ curve of the NPEB portfolio
is also obtained by Monte Carlo simulations with 500 runs, by using
different values of $\lambda>0$ in a grid. This curve is relatively
close to Markowitz's efficient frontier among the $(\sigma, \mu)$
curves of various portfolios that do not assume knowledge of $\bmu$
and $\bSigma$, as shown in Figure \ref{effi}. For the
covariance-shrinkage portfolio,
we use a constant correlation model for $\widehat{F}$ in (\ref{eq24}), which
can be implemented by their software available at \href{http://www.ledoit.net}{www.ledoit.net}.
Note that Markowitz's efficient frontier has $\mu$ values ranging
from 2.0 to 3.47, which is the largest component of $\bmu$ in
Freq 1. The $(\sigma, \mu)$ curve of NPEB lies below the
efficient frontier, and further below are the $(\sigma, \mu)$ curves
of Michaud's, covariance-shrinkage and plug-in portfolios, in decreasing
order. These $(\sigma, \mu)$ curves are what Broadie (\citeyear{Broadie1993}) calls the
\textit{actual frontiers}.

The highest values 3.22, 3.22 and 3.16 of $\mu$ for the plug-in,
covariance-shrinkage
and Michaud's portfolios in Figure \ref{effi} are attained with a
target value
$\mu_* = 3.47$, and the corresponding values of $\sigma$ are 1.54, 1.54
and 3.16, respectively. Note that without short selling, the constraint
${\mathbf w}^T \widehat{\bmu}=\mu_*$ used in these portfolios cannot hold
if $\max_{1\le i \le4} \widehat{\mu}_i <\mu_*$.
We therefore need a default option, such as replacing
$\mu_*$ by $\min(\mu_*, \max_{1\le i \le4}\widehat{\mu}_i)$, to implement
the optimization procedures for these portfolios. In contrast, the NPEB
portfolio can always be implemented for any given value of $\lambda$.
In particular, for $\lambda=0.001$, the NPEB portfolio has $\mu=3.470$
and $\sigma=1.691$.

\section{Connecting theory to practice}\label{sec5}

While Section \ref{sec4} has considered practical implementation of the theory
in Section \ref{sec3}, we develop the methodology further in this section to
connect the basic theory to practice.

\subsection{\texorpdfstring{The information ratios and choice of $\lambda$}{The information ratios and choice of lambda}}
\label{sec51}

As pointed out in Section~\ref{sec1}, the~$\lambda$ in Section \ref{sec3} is related to
how risk-averse one is when one tries to maximize the expected utility
of a portfolio. It represents a penalty on the risk that is
measured by the variance of the portfolio's return.
In practice, it may be difficult to specify an investor's risk
aversion parameter $\lambda$ that is needed in the
theory in Section \ref{sec31}. A~commonly used performance measure
of a portfolio's performance is the information ratio
$(\mu-\mu_0)/\sigma_e$, which is the excess return per unit of risk;
the excess is measured by $\mu-\mu_0$, where $\mu_0=E(r_0)$,~$r_0$ is
the return of the benchmark investment and $\sigma_e^2$ is the
variance of the excess return. We can regard $\lambda$ as a tuning
parameter, and choose it to maximize the information ratio by
modifying the NPEB procedure in Section~\ref{sec32},
where the bootstrap estimate of $E_{\bmu, \bSigma}[ {\mathbf
w}^T(\eta)
{\mathbf r} ] - \lambda\operatorname{Var}_{\bmu, \bSigma}[ {\mathbf
w}^T(\eta)
{\mathbf r} ] $ is used to find the portfolio weight
${\mathbf w}_{\lambda}$ that solves the optimization problem~(\ref{eq33}).
Specifically, we use the bootstrap estimate of the information ratio
\begin{equation}\label{eq51}
E_{\bmu, \bSigma} ( {\mathbf w}_\lambda{\mathbf r} - r_0) /
\sqrt{\operatorname{Var}_{\bmu, \bSigma}({\mathbf w}_\lambda^T {\mathbf r} - r_0)}
\end{equation}
 of ${\mathbf w}_{\lambda}$, and maximize the estimated information
ratios over
$\lambda$ in a grid that will be illustrated in Section \ref{sec6}.

\subsection{Dimension reduction when $m$ is not small relative to
$n$}\label{sec52}

Another~sta\-tistical issue encountered in practice is the large number $m$
of assets relative to the number $n$ of past periods in the training sample,
making it difficult to estimate $\bmu$ and $\bSigma$ satisfactorily.
Using factor models that are related to domain knowledge as in Section
\ref{sec21}
helps reduce the number of parameters to be estimated in an empirical
Bayes approach.

%
\begin{table}[b]
\vspace*{-6pt}
\tabcolsep=5pt
\caption{Rewards of four portfolios formed from $m=6$ assets}\label{table:6assets}%
\vspace*{-5pt}
\begin{tabular*}{\textwidth}{@{\extracolsep{\fill}}lccccccc@{}}
\hline
$\bolds\lambda$ & $\bolds{(}\bmu\bolds{,} \bSigma\bolds{)}$ & \textbf{Bayes} & \textbf{Plug-in} & \textbf{Oracle} & \textbf{NPEB} \\
\hline
\phantom{0}1 & Bayes& 0.0325 (2.55e$-$5) & 0.0318 (2.62e$-$6) & 0.0331
(2.42e$-$5) & 0.0325 (2.53e$-$5) \\
&Freq 1& 0.0284 (1.59e$-$5) & 0.0277 (1.31e$-$5) & 0.0296\phantom{ (2.62e$-$6)} & 0.0285
(1.62e$-$5) \\
&Freq 2& 0.0292 (8.30e$-$6) & 0.0280 (7.95e$-$6) & 0.0296\phantom{ (2.62e$-$6)} & 0.0292
(8.29e$-$6) \\
&Freq 3& 0.0283 (1.00e$-$5) & 0.0282 (9.11e$-$6) & 0.0300\phantom{ (2.62e$-$6)} & 0.0283
(1.05e$-$5)
\\ [3pt]
\phantom{0}5 & Bayes& 0.0255 (2.46e$-$5) & 0.0183 (1.44e$-$5) & 0.0263
(2.05e$-$5) & 0.0254 (2.45e$-$5) \\
&Freq 1& 0.0236 (1.99e$-$5) & 0.0149 (6.48e$-$6) & 0.0250\phantom{ (2.05e$-$6)} & 0.0237
(2.17e$-$5) \\
&Freq 2& 0.0241 (9.34e$-$6) & 0.0166 (3.61e$-$6) & 0.0246\phantom{ (2.05e$-$6)} & 0.0243
(8.95e$-$6) \\
&Freq 3& 0.0189 (2.09e$-$5) & 0.0138 (1.45e$-$5) & 0.0219\phantom{ (2.05e$-$6)} & 0.0208
(2.32e$-$5)
\\ [3pt]
10 & Bayes& 0.0171 (2.63e$-$5) & 0.0039 (1.57e$-$5) & 0.0180
(2.20e$-$5) & 0.0171 (2.72e$-$5) \\
&Freq 1&0.0174 (2.06e$-$5) & 0.0042 (5.19e$-$6) & 0.0193\phantom{ (2.20e$-$6)} & 0.0177
(2.42e$-$5) \\
&Freq 2& 0.0177 (1.12e$-$5) & 0.0052 (6.34e$-$6) & 0.0184\phantom{ (2.20e$-$6)} & 0.0180
(1.10e$-$5) \\
&Freq 3& 0.0089 (2.79e$-$5) & 0.0024 (1.33e$-$5) & 0.0120\phantom{ (2.20e$-$6)} & 0.0094
(4.65e$-$5)\\
\hline
\end{tabular*}
\end{table}

An obvious way of dimension reduction when there is no short selling
is to exclude assets with markedly inferior information ratios from
consideration. The only potential advantage of including them
in the portfolio is that they may be able to reduce the portfolio
variance if
they are negatively correlated with the ``superior'' assets. However, since
the correlations are unknown, such advantage is unlikely when they are not
estimated well enough.
Suppose we include in the simulation study of Section \ref{sec42}
two more assets so that all asset returns are jointly normal.
The additional hyperparameters of the normal and inverted Wishart prior
distribution (\ref{eq22}) are $\bnu_5= -0.014$, $\bnu_6=-0.064$,
$\bPsi_{55}=2.02$, $\bPsi_{66}=10.32$, $\bPsi_{56}=0.90$,
$\bPsi_{15}=-0.17$, $\bPsi_{25}=-0.03$, $\bPsi_{35}=-0.91$, $\bPsi_{45}=
-0.33$, $\bPsi_{16}=-3.40$, $\bPsi_{26}=-3.99$, $\bPsi_{36}=-0.08$ and
$\bPsi_{46}=-3.58$. As in Section \ref{sec42}, we consider four scenarios for
the case of $n=8$ without short selling, the first of which assumes this
prior distribution and studies the Bayesian reward for $\lambda=1, 5$
and 10.
Table \ref{table:6assets} shows the rewards for the four rules in
Section \ref{sec42},
and each result is based on 500 simulations. Note that
the value of the reward function does not show significant change with
the inclusion of two additional stocks, which have negative correlations
with the four stocks in Section \ref{sec42} but have low information
ratios. This shows that excluding stocks with markedly inferior
information ratios when there is no short selling can reduce $m$
substantially in practice. In Section~\ref{sec6} we describe another way of
choosing stocks from a universe of available stocks to reduce $m$.

\subsection{Extension to time series models of returns}\label{sec53}

An important assumption in the modification of Markowitz's theory in
Section \ref{sec32} is that ${\mathbf r}_t$ are i.i.d. with mean $\bmu$ and covariance
matrix $\bSigma$. Diagnostic checks of the extent to which this
assumption is violated should be carried out in practice. The
stochastic optimization theory in Section \ref{sec31} does not actually need this
assumption and only requires the posterior mean and second moment
matrix of the return vector for the next period in~(\ref{eq34}).
Therefore, one can modify the ``working i.i.d.
model'' accordingly when the diagnostic checks reveal such modifications
are needed.

A simple method to introduce such modification is to use a stochastic
regression model of the form
\begin{equation}\label{eq52}
r_{it} = \bbeta_i^T {\mathbf x}_{i,t-1} + \epsilon_{it},
\end{equation}
 where the components of ${\mathbf x}_{i,t-1}$ include 1, factor variables
such as the return of a~market portfolio like S\&P500 at time $t-1$,
and lagged
variables $r_{i,t-1}, r_{i,t-2},\ldots.$ The basic idea underlying (\ref{eq52}) is
to introduce covariates (including lagged variables to account for
time series effects) so that the errors $\epsilon_{it}$ can be regarded
as i.i.d., as in the working i.i.d. model. The regression parameter
$\bbeta_i$ can be estimated by the method of moments, which is equivalent
to least squares. We can also include heteroskedasticity
by assuming that $\epsilon_{it}=s_{i,t-1}(\bgamma_i)
z_{it}$, where $z_{it}$ are i.i.d. with mean 0 and variance 1,~%
$\bgamma_i$ is a parameter vector which can be estimated by
maximum likelihood or generalized method of moments, and $s_{i,t-1}$ is a
given function that depends on $r_{i,t-1}, r_{i,t-2}, \ldots.$ A
well-known example is the $\operatorname{GARCH}(1,1)$ model
\begin{equation}\label{eq53}
\epsilon_{it}=s_{i,t-1} z_{it},\qquad
s_{i,t-1}^2 = \omega_i + a_i s_{i,t-2}^2 + b_i r_{i,t-1}^2
\end{equation}
for which $\bgamma_i = (\omega_i, a_i, b_i)$.

Consider the stochastic regression model (\ref{eq52}).
As noted in Section \ref{sec32}, a~key ingredient in the optimal weight vector that
solves the optimization problem (\ref{eq31}) is $(\bmu_n, {\mathbf V}_n)$, where
$\bmu_n = E({\mathbf r}_{n+1} |{\mathcal R}_n)$ and ${\mathbf V}_n = E({\mathbf r}_{n+1}
{\mathbf r}_{n+1}^T |\break {\mathcal R}_n)$. Instead of the classical model of i.i.d.
returns, one can combine domain knowledge of the $m$ assets with time
series modeling to obtain better predictors of future returns via $\bmu_n$
and ${\mathbf V}_n$. The regressors ${\mathbf x}_{i,t-1}$ in (\ref{eq52}) can be
chosen to
build a combined substantive--empirical model for prediction; see
Section 7.5 of Lai and Xing~(\citeyear{LX2008}). Since the model (\ref{eq52}) is intended
to produce i.i.d. $\bepsilon_t = (\epsilon_{1t}, \ldots,
\epsilon_{mt})^T$, or i.i.d. ${\mathbf z}_t= (z_{1t}, \dots, z_{mt})^T$
after adjusting for conditional heteroskedasticity as in (\ref{eq53}), we can
still use the NPEB approach to determine the optimal weight
vector, bootstrapping from the estimated common distribution of
$\bepsilon_t$ (or ${\mathbf z}_t$). Note that (\ref{eq52}) and (\ref{eq53}) models the asset
returns separately, instead
of jointly in a multivariate regression or multivariate $\operatorname{GARCH}$ model
which has too many parameters to estimate.
While the vectors $\bepsilon_t$ (or ${\mathbf z}_t$) are assumed to be i.i.d.,
(\ref{eq52}) [or (\ref{eq53})] does not assume their components to be uncorrelated
since it treats the components separately rather than jointly. The
conditional cross-sectional covariance between the returns of assets $i$
and $j$ given ${\mathcal R}_n$ is given by
\begin{equation}\label{sec54}
\operatorname{Cov}(r_{i,n+1}, r_{j,n+1} | {\mathcal R}_n) = s_{i,n}(\bgamma_i)
s_{j,n}(\bgamma_j) \operatorname{Cov}( z_{i,n+1}, z_{j,n+1} | {\mathcal
R}_n)
\end{equation}
 for the model (\ref{eq52}) and (\ref{eq53}). Note that (\ref{eq53}) determines $s_{i,n}^2$
recursively from~${\mathcal R}_n$, and that ${\mathbf z}_{n+1}$ is independent of
${\mathcal R}_n$ and, therefore, its covariance matrix can be
consistently estimated from the residuals $\widehat{\mathbf z}_t$. Under
(\ref{eq52}) and (\ref{eq53}), the NPEB approach uses the following formulas for $\bmu_n$
and ${\mathbf V}_n$ in~(\ref{eq35}):
\begin{equation}\label{sec55}
\hspace*{25pt}\bmu_n = (\widehat{\bbeta}_1^T {\mathbf x}_{1,n}, \dots,
\widehat{\bbeta}_m^T {\mathbf x}_{m,n})^T,\qquad
{\mathbf V}_n = \bmu_n \bmu_n^T + ( \widehat{s}_{i,n}
\widehat{s}_{j,n} \widehat{\sigma}_{ij} )_{1\le i, j \le n},
\end{equation}
in which $\widehat{\bbeta}_i$ is the least squares estimate of
$\bbeta_i$, and $\widehat{s}_{l,n}$ and $\widehat{\sigma}_{ij}$ are the
usual estimates of $s_{l,n}$ and $\operatorname{Cov}(z_{i,1}, z_{j,1})$ based on
${\mathcal
R}_n$. Further discussion of time series modeling for implementing the
optimal portfolio in Section \ref{sec3} will be given in Sections \ref{sec62} and \ref{sec7}.
\vspace*{-3pt}

\section{An empirical study}\label{sec6}

In this section we describe an empirical study of the out-of-sample
performance of the proposed approach and other methods for
mean--variance portfolio optimization when the means
and covariances of the underlying asset returns are unknown. The
study uses monthly stock market data from January 1985 to December
2009, which are obtained from the Center for Research in Security
Prices (CRSP) database, and evaluates out-of-sample performance of
different portfolios of these stocks for each month after the first
ten years (120~months) of this period to accumulate training data. The
CRSP database can be accessed through the Wharton Research Data Services
at the University of Pennsylvania
(\url{http://wrds.wharton.upenn.edu}). Following
Ledoit and Wolf (\citeyear{LW2004}), at the beginning of month $t$, with $t$
varying from January 1995 to December 2009, we select $m=50$ stocks
with the largest market values among those that have no missing
monthly prices in the previous 120 months, which are used as the
training sample. The portfolios for month $t$ to be considered are
formed from these $m$ stocks.

Note that this period contains highly volatile times in the stock
market, such as around ``Black Monday'' in 1987, the Internet bubble
burst and the September 11 terrorist attacks in 2001, and the ``Great
Recession'' that began in 2007 with the default and other difficulties
of subprime mortgage loans. We use sliding windows of $n=120$ months
of training data to construct portfolios of the stocks for the subsequent
month. In contrast to the Black--Litterman approach described in
Section \ref{sec22}, the portfolio construction is based solely on these data
and uses no other information about the stocks and their associated
firms, since the purpose of the empirical study is to illustrate the
basic statistical aspects of the proposed method and to compare it
with other statistical methods for implementing Markowitz's
mean--variance portfolio optimization theory. Moreover, for a fair
comparison, we do not assume any prior distribution as in the Bayes
approach, and only use NPEB in this study.

Performance of a portfolio is measured by the excess returns $e_t$
over a~benchmark portfolio. As $t$ varies over the monthly test periods
from January 1995 to December 2009, we can (i) add up the realized
excess returns to give the \textit{cumulative realized excess return}
$\sum_{l=1}^t e_l$ up to time $t$, and (ii)~use the average realized
excess return and the standard deviation to evaluate the
\textit{realized information ratio} $\sqrt{12}\bar{e}/ s_e$, where
$\bar{e}$ is the sample average of the monthly excess returns and
$s_e$ is the corresponding sample standard deviation, using
$\sqrt{12}$ to annualize the ratio as in Ledoit and Wolf
(\citeyear{LW2004}). Noting that the realized information ratio is a summary
statistic of the monthly excess returns in the 180 test periods, we find
it more informative to supplement this commonly used measure of
investment performance with the time series plot of cumulative realized
excess returns, from which the realized excess returns $e_t$ can be
retrieved by differencing.

We use two ways to construct the benchmark portfolio. The first
follows that of Ledoit and Wolf (\citeyear{LW2004}), who propose to mimic how an
active portfolio manager chooses the benchmark to define excess
returns. It is described in Section \ref{sec61}. The second simply uses the
S\&P500 Index as the benchmark portfolio and Section \ref{sec63} considers
this case. Section \ref{sec62} compares the time series of the returns of
these two benchmark portfolios and explains why we choose to use the
S\&P500 Index as the benchmark portfolio in conjunction with the time
series model (\ref{eq52}) and (\ref{eq53}) for the excess returns in
Sec\-tion~\ref{sec63}.\vspace*{-3pt}

\subsection{Active portfolios and associated optimization
problems}\label{sec61}
In this section the benchmark portfolio consists of the $m=50$
stocks chosen at the beginning of each test period and
weights them by their market values. Let~${\mathbf w}_{B}$ denote the
weight of this value-weighted benchmark and ${\mathbf w}$ the weight of
a~given portfolio. The difference $\widetilde{\mathbf w}={\mathbf w} - {\mathbf
w}_{B}$ satisfies $\widetilde{\mathbf w}^T {\mathbf1} = 0$. An active portfolio
manager would choose ${\mathbf w}$ that solves the following optimization
problem instead of (\ref{eq11}):\vspace*{-3pt}
\begin{eqnarray}\label{eq61}
{\mathbf w}_{\mathrm{active}} & = &{\mathbf w}_B+\arg\min_{\widetilde{\mathbf w}}
\widetilde{\mathbf w}^T \bSigma\widetilde{\mathbf w}\quad \mbox{subject to}\quad
\widetilde{\mathbf w}^T \bmu = \widetilde{\mu}_*, \nonumber\\ [-10pt]\\ [-10pt]
 \widetilde{\mathbf w}^T {\mathbf1}&=&0 \mbox{ and }
\widetilde{\mathbf w} \in{\mathcal C},\nonumber\vspace*{-3pt}
\end{eqnarray}
in which ${\mathcal C}$ represents additional constraints for the
manager, $\bSigma$ is the covariance matrix of stock returns and
$\widetilde{\mu}_*$ is the target excess return over the
value-weighted benchmark.
The portfolio defined by ${\mathbf w}_{\mathrm{active}}$ is
called 
an \textit{active portfolio}. Since $\bmu$ and $\bSigma$ are typically
unknown, putting a prior distribution on them in (\ref{eq61}) leads to the
following modification of (\ref{eq31}):
\begin{equation}\label{eq62}
\max \{ E(\widetilde{\mathbf w}^T {\mathbf r}_{n+1}) - \lambda
\operatorname{Var}(\widetilde{\mathbf w}^T {\mathbf r}_{n+1}) \}\quad
\mbox{subject to}\quad \widetilde{\mathbf w}^T {\mathbf1}=0.
\end{equation}
This optimization problem can be solved by the same method as that
introduced in Section \ref{sec3}.

%
\begin{table}
\caption{Means and standard deviations (in parentheses) of the
annualized realized excess returns over the value-based benchmark}
\label{emp_table1}
\begin{tabular*}{\textwidth}{@{\extracolsep{\fill}}lcccc@{}}
\hline
$\bolds{\widetilde{\mu}_*}$ & \textbf{0.01} &\textbf{0.015} & \textbf{0.02} & \textbf{0.03} \\
$\bolds\lambda$ & $\bolds{2^2}$ & \textbf{2} & $\bolds{2^{-1}}$ & $\bolds{2^{-2}}$ \\
\hline
\multicolumn{5}{@{}l}{(a) All test periods by re-defining portfolios in
some periods} \\
\quad Plug-in & 0.001 (4.7e$-$3) & 0.002 (7.3e$-$3) & 0.003 (9.6e$-$3) & 0.007
(1.4e$-$2)\\
\quad Shrink & 0.003 (4.3e$-$3) & 0.004 (6.6e$-$3) & 0.006 (8.8e$-$3) & 0.011
(1.3e$-$2)\\
\quad Boot & 0.001 (2.5e$-$3) & 0.001 (3.8e$-$3) & 0.001 (5.1e$-$3) & 0.003
(7.3e$-$3)\\ 
\quad NPEB & 0.029 (1.2e$-$1) & 0.046 (1.3e$-$1) & 0.053 (1.5e$-$1) & 0.056
(1.6e$-$1)\\ [3pt] 
\multicolumn{5}{@{}l}{(b) Test periods in which all portfolios are well
defined}\\
\quad Plug-in &0.002 (6.6e$-$3) & 0.004 (1.0e$-$2) & 0.006 (1.4e$-$2) & 0.014
(1.9e$-$2)\\ 
\quad Shrink & 0.005 (5.9e$-$3) & 0.008 (9.0e$-$3) & 0.012 (1.2e$-$2) & 0.021
(1.8e$-$2)\\ 
\quad Boot &0.001 (3.5e$-$3) & 0.003 (5.3e$-$3) & 0.003 (7.1e$-$3) & 0.006
(1.0e$-$2)\\ 
\quad NPEB & 0.282 (9.3e$-$2) & 0.367 (1.1e$-$1) & 0.438 (1.1e$-$1) & 0.460
(1.1e$-$2)\\ 
\hline
\end{tabular*}
\end{table}

Following Ledoit and Wolf (\citeyear{LW2004}), we choose the constraint set ${\mathcal
C}$ such that the portfolio is long only and the total position in any
stock cannot exceed an upper bound $c$, that is, ${\mathcal C}=\{
\widetilde{\mathbf w}\dvtx -{\mathbf w}_B \le\widetilde{\mathbf w} \le c {\mathbf
1}-{\mathbf
w}_B \}$, with $c=0.1$. We use quadratic programming to solve the
optimization problem~(\ref{eq61}) in which $\bmu$ and $\bSigma$ are
replaced, for the plug-in active portfolio, by their sample estimates
based on
the training sample in the past 120 months. The covariance-shrinkage
active portfolio uses a shrinkage estimator of $\bSigma$ instead,
shrinking toward a patterned matrix that assumes all pairwise
correlations to be equal [Ledoit and Wolf (\citeyear{LW2003})]. Similarly, we can
extend Section \ref{sec23} to obtain a resampled active portfolio, and also
extend the NPEB approach in Section \ref{sec4} to construct the corresponding
NPEB active portfolio. Table~\ref{emp_table1} summarizes the realized
information ratio $\sqrt{12} \bar{e} / s_e$ for different values of
annualized target excess returns $\widetilde{\mu}_*$ and ``matching''
values of $\lambda$ whose choice is described below.

We first note that specified target returns $\widetilde{\mu}_*$ may be
vacuous for the plug-in, covariance-shrinkage (abbreviated ``shrink''
in Table \ref{emp_table1}) and resampled (abbreviated ``boot'' for
bootstrapping) active portfolios in a given test period. For
$\widetilde{\mu}_* = 0.01, 0.015, 0.02, 0.03$, there are 92, 91, 91
and 80 test periods, respectively, for which (\ref{eq61}) has solutions when
$\bSigma$ is replaced by either the sample covariance matrix or the
Ledoit--Wolf shrinkage estimator of the training data from the previous
120 months. Higher levels of target returns result in even fewer of the
180 test periods for which (\ref{eq61}) has solutions. On the other hand,
values of $\widetilde{\mu}_*$ that are lower than 1\% may be of little
practical interest to active portfolio managers. When (\ref{eq61}) does not
have a solution to provide a portfolio of a specified type for a test
period, we use the value-weighted benchmark as the portfolio for the
test period. Table \ref{emp_table1}(a) gives the actual (annualized)
mean realized excess returns $12\bar{e}$ to show the extent to which
they match the target value $\widetilde{\mu}_*$, and also the
corresponding annualized standard deviations $\sqrt{12}s_e$, over the
180 test periods for the plug-in, covariance-shrinkage and resampled
active portfolios constructed with the above modification.
These numbers are very small, showing that the three portfolios differ
little from the benchmark portfolio, so the realized information
ratios that range from 0.24 to 0.83 for these active portfolios can be
quite misleading if the actual mean excess returns are not taken into
consideration.\looseness=-1

We have also tried another default option that uses 10 stocks with the
largest mean returns (among the 50 selected stocks) over the training
period and puts equal weights to these 10 stocks to form a portfolio
for the ensuing test period for which (\ref{eq61}) does not have a
solution. The mean realized excess returns $12\bar{e}$ when this
default option is used are all negative (between $-$17.4\% and $-$16.3\%),
while $\widetilde{\mu}_*$ ranges from 1\% from 3\%.
Table \ref{emp_table1}(a) also gives the means and standard deviations
of the annualized realized excess returns of the NPEB active portfolio
for four values of $\lambda$ that are
chosen so that the mean realized excess returns roughly match the
values of $\widetilde{\mu}_*$ over a grid of the form $\lambda= 2^j$
$(-2 \le j \le2)$ that we have tried. Note that NPEB has considerably
larger mean excess returns than the other three portfolios.

Table \ref{emp_table1}(b) restricts only to the 80--92 test
periods in which the plug-in, covariance-shrinkage
and resampled active portfolios are all well defined by~(\ref{eq61}) for
$\widetilde{\mu}_* = 0.01, 0.015, 0.02$ and 0.03. The mean excess
returns of the plug-in, covariance-shrinkage and resampled portfolios
are still very small, while those of NPEB are much larger. The
realized information ratios of NPEB range from 3.015 to 3.954, while
those of the other three portfolios range from 0.335 to 1.214 when we
restrict to these test periods.

\subsection{Value-weighted portfolio versus S\&P500 Index and time series
effects}\label{sec62}

The results for the plug-in and covariance-shrinkage portfolios in
Table \ref{emp_table1} are markedly different from those of Ledoit
and Wolf (\citeyear{LW2004}) covering a different period (February 1983--December 2002).
This suggests that the stock returns cannot be
approximated by the assumed i.i.d. model underlying these methods. In
Section~\ref{sec53} we have extended the NPEB approach to~a~ve\-ry flexible time
series model~(\ref{eq52}) and (\ref{eq53}) of the stock returns~$r_{it}$. The stochastic
regression model~(\ref{eq52}) can incorporate important time-varying
predictors in~${\mathbf x}_{it}$ for the $i$th stock's performance at time
$t$, while the $\operatorname{GARCH}$ mo\-del~(\ref{eq53}) for the random disturbances
$\epsilon_{it}$ in (\ref{eq52}) can incorporate dynamic features of the
stock's idosyncratic variability. It seems that a regressor such as
the return $u_t$ of S\&P500 Index should be included in ${\mathbf x}_{it}$
to take advantage of the co-movements of $r_{it}$ and $u_t$. However,
since $u_t$ is not observed at time $t$, one may need to have good
predictors of $u_t$ which should consist not only of the past S\&P500
returns but also macroeconomic variables. Of course, stock-specific
information such as the firm's earnings performance and forecast and
its sector's economic outlook should also be considered. This means
that fundamental analysis, as carried out by professional
stock analysts and economists in investment banks, should be
incorporated into the model (\ref{eq52}). Since this is clearly beyond the
scope of the present empirical study whose purpose is to illustrate our
new
statistical approach to the Markowitz optimization enigma, we shall
focus on simple models to demonstrate the benefit of building good
models for ${\mathbf r}_{t+1}$ in our stochastic optimization approach.

%
\begin{figure}

\includegraphics{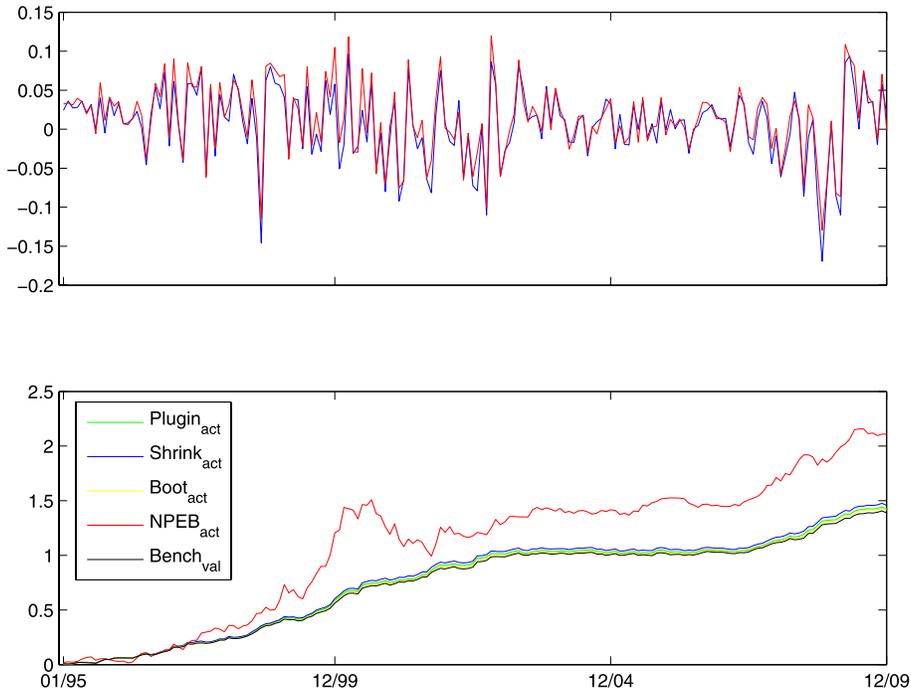}

\caption{Time series plots of the monthly returns \textup{(Top)} of the S\&P500 Index
and the value-weighted benchmark, and cumulative excess returns
\textup{(Bottom)}
of the value-weighted benchmark (abbreviated by Bench$_{\mathrm{val}}$) and
for active portfolios (abbreviated by the subscript ``act'') over the
S\&P500 Index.}\label{fig_bench}
\end{figure}

%
\begin{figure}

\includegraphics{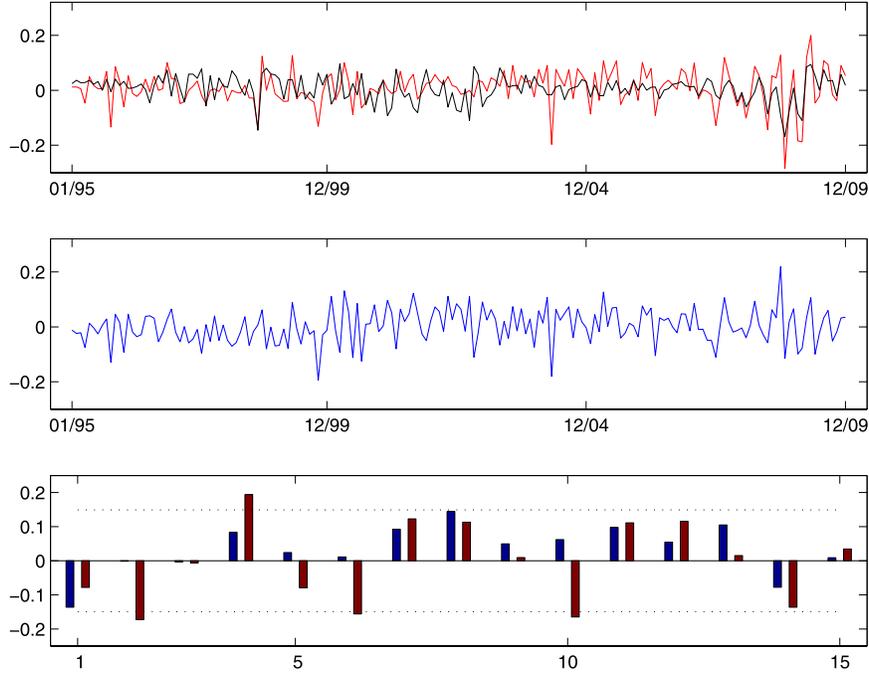}

\caption{Comparison of returns and excess returns. \textup{Top panel:}
returns of S\&P500 Index (black) and SHW (red). \textup{Middle
panel:} excess returns (blue) of SHW. \textup{Bottom panel:}
Autocorrelations of returns (red) and excess returns (blue) of
SHW; the dotted lines represent rejection boundaries of
5\%-level tests of zero autocorrelation at indicated
lag.}\label{fig_comp}
\end{figure}

In this connection, we first compare the S\&P500 Index with the
value-weighted portfolio, which is the benchmark portfolio in Section
\ref{sec61}. The top panel of Figure~\ref{fig_bench} gives the time series plots
of the monthly returns (which are not annualized) of both portfolios
during the test period. The S\&P500 Index has mean 0.006 and
standard deviation 0.046 in this period, while the mean of
the value-weighted portfolio is 0.0137 and its standard deviation is
0.045. The bottom panel of Figure \ref{fig_bench} plots the time series
of cumulative realized excess returns $\sum_{l=1}^t e_l$ over the
S\&P500 Index, for the value-weighted portfolio and also for the
four active portfolios in Table \ref{emp_table1}(a) under the column
$\mu_*=0.015$ and $\lambda= 2$, during the test period (January 1995--December 2009). Unlike NPEB$_{\mathrm{act}}$, the cumulative realized
excess returns of the other three active portfolios differ little from the
value-weighted portfolio, as shown by the figure.

In view of the structural changes in the economy and the financial
markets during this period, it appears difficult to find simple time
series models that can reflect the inherent nonstationarity. If we use
the S\&P500 Index $u_t$ as an alternative benchmark to the
value-weighted portfolio used in Section \ref{sec61}, the excess returns
$e_{it} = r_{it}-u_t$ may be able to exploit the co-movements of
$r_{it}$ and $u_t$ to remove their common nonstationarity due to changes
in macroeconomic variables. As an illustration, the top panel of
Figure~\ref{fig_comp} gives the time series plots of returns of
Sherwin--Williams Co. (SHW) and of the S\&P500 Index during this
period, and the middle panel gives the time series plot of the excess
returns. The Ljung--Box test, which involves autocorrelations of lags up
to 20 months, has $p$-value 0.001 for the monthly returns of
SHW and 0.267 for the excess returns, and therefore rejects the
i.i.d. assumption for the actual but not the excess returns; see
Section~5.1 of Lai and Xing (\citeyear{LX2008}). This is also shown graphically by
the autocorrelation functions in the bottom panel of Figure
\ref{fig_comp}.

\subsection{Using the S\&P500 Index as benchmark portfolio and time series
models of excess returns}\label{sec63}
The preceding section shows that using the S\&P500 Index as the
benchmark portfolio has certain advantages over the value-weighted
portfolio. In this section we consider the excess returns $e_{it}$ over
the S\&P500 Index $u_t$, which we use as the benchmark portfolio, and
fit relatively simple time series models to the training sample to
predict the mean and volatility of $e_{it}$ for the test period. Instead
of forming active portfolios as in Section \ref{sec61}, we follow traditional
portfolio theory as described in Sections \ref{sec1}--\ref{sec3}. Note that this theory
assumes the constraint $\sum_{i=1}^m w_i=1$ and, therefore,
\begin{equation}\label{eq63}
\sum_{i=1}^m w_i r_{it} - u_t = \sum_{i=1}^m w_i (r_{it}-u_t) =
\sum_{i=1}^m w_i e_{it},
\end{equation}
whereas active portfolio optimization considers weights
$\widetilde{w}_i = w_i - w_{i,B}$ that satisfy the constraint
$\sum_{i=1}^m \widetilde{w}_i = 0$. In view of (\ref{eq63}), when the
objective is to maximize the mean return of the portfolio subject to a
constraint on the volatility of the excess return over the benchmark
(which is related to achieving an optimal information ratio), we can
replace the returns $r_{it}$ by the excess returns $e_{it}$ in the
portfolio optimization problem (\ref{eq11}) or (\ref{eq31}). As explained in the
second paragraph of Section~\ref{sec62}, $e_{it}$ can be modeled by simpler
stationary time series models than~$r_{it}$.

The simplest time series model to try is the $\operatorname{AR}(1)$ model $e_{it}=\alpha
_i +
\gamma_i e_{i,t-1} + \epsilon_{it}$. Assuming this time series model
for the excess returns, we can apply the NPEB procedure in Section
\ref{sec53}
to the training sample and thereby obtain the NPEB$_{\operatorname{AR}}$ portfolio
for the test sample. The $\operatorname{AR}(1)$ model uses ${\mathbf x}_{i,t-1} = (1,
e_{i,t-1})^T$ as the predictor in a linear regression
model for $e_{i,t}$. To improve prediction performance, one can include
additional predictor variables, for example, the return $u_{t-1}$ of
the S\&P500
Index in the preceding period. Assuming the stochastic regression model
$e_{i,t}=(1, e_{i,t-1}, u_{t-1}) \bbeta_i + \epsilon_{i,t}$, and the
$\operatorname{GARCH}(1,1)$ model (\ref{eq53}) for $\epsilon_{i,t}$, we can apply the NPEB
procedure to the training sample and thereby form the NPEB$_{\mathrm{SRG}}$
portfolio for the test sample.

\begin{figure}[t]

\includegraphics{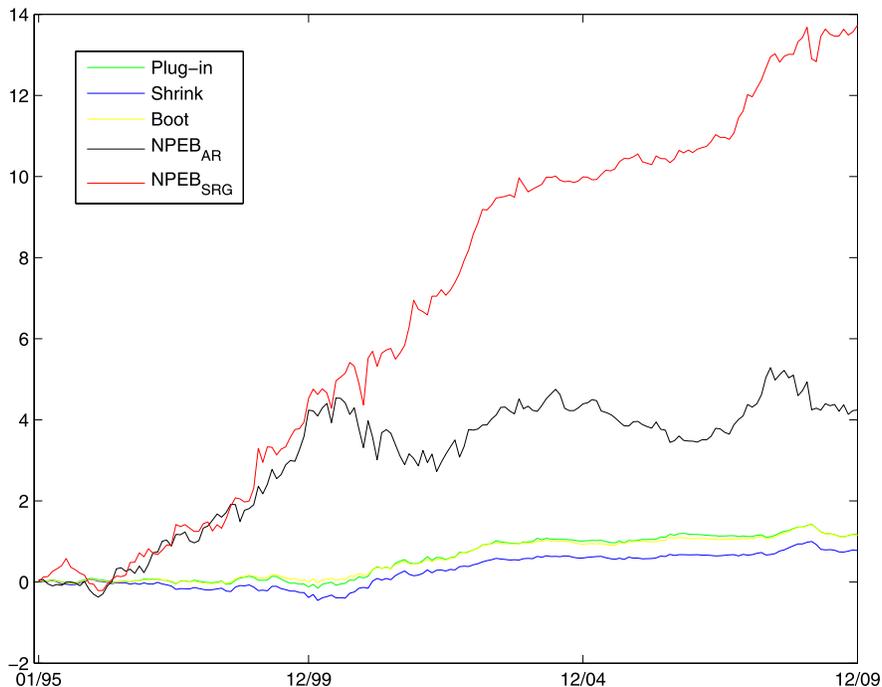}

\caption{Realized cumulative excess returns over the S\&P500 Index.}
\label{fig_cumsum}
\end{figure}

Instead of taking long-only positions (i.e., $w_i \ge0$ for all $i$),
we also allow short selling, with the constraint $w_i \ge-0.05$ for all
$i$, to construct the following portfolios in this section. For the
plug-in, covariance-shrinkage and resampled portfolios, which we
abbreviate as in Figure \ref{fig_bench} but without the subscript
``act'' (for active), we use the annualized target return $\mu_*=0.015,
0.02, 0.03$, for which the problem (\ref{eq11}) can be solved for all 180
test periods under the weight constraint; note that
we use the mean return instead of the mean excess return as the target
$\mu_*$. For the NPEB$_{\operatorname{AR}}$ and NPEB$_{\mathrm{SRG}}$ portfolios, we use
the training sample as in Section \ref{sec51} to choose $\lambda$ by maximizing
the information ratio over the grid $\lambda\in\{ 2^i\dvtx i=-3, -2,
\dots, 6 \}$. Figure \ref{fig_cumsum} plots the time series of
cumulative realized excess returns over the S\&P500 Index during the
test period of 180 months, for Plug-in, Shrink and Boot with $\mu_* =
0.015$ and for NPEB$_{\operatorname{AR}}$ and NPEB$_{\mathrm{SRG}}$. Table~\ref{emp_table2} gives the annualized realized information ratios, with
the S\&P500 Index as the benchmark portfolio. The table also considers
cases $\mu_*=0.02, 0.03$, and further abbreviates Plug-in, Shrink and
Boot by P, S, B, respectively.

%

%
\begin{table}
\tabcolsep=0pt
\caption{Realized information ratios and average realized excess
returns (in square brackets)
with respect to the S\&P500 Index}\label{emp_table2}
\begin{tabular*}{\textwidth}{@{\extracolsep{\fill}}lcccccccccc@{}}
\hline
\multicolumn{3}{@{}c}{$\bolds{\mu_*=0.015}$} & \multicolumn{3}{c}{$\bolds{\mu_*=0.02}$}
&\multicolumn{3}{c}{$\bolds{\mu_*=0.03}$} & \multicolumn{2}{c@{}}{\textbf{NPEB}} \\ [-7pt]
\multicolumn{3}{@{}c}{\hrulefill} & \multicolumn{3}{c}{\hrulefill}
&\multicolumn{3}{c}{\hrulefill} & \multicolumn{2}{c@{}}{\hrulefill}\\
\textbf{P} & \textbf{S} & \textbf{B} & \textbf{P} & \textbf{S} & \textbf{B} & \textbf{P}
& \textbf{S} & \textbf{B} & $\bolds{\operatorname{AR}}$ & \textbf{SRG} \\
\hline
0.527 & 0.352 & 0.618 & 0.532 & 0.353 & 0.629 & 0.538 & 0.354 &
0.625 & 0.370 & 1.169 \\
\mbox{[0.078]} & \mbox{[0.052]} & \mbox{[0.077]} & \mbox{[0.078]} & \mbox{[0.051]} & \mbox{[0.078]} &
\mbox{[0.076]} & \mbox{[0.050]} & \mbox{[0.077]} & \mbox{[0.283]} & \mbox{[0.915]} \\ \hline
\end{tabular*}
\end{table}

\subsection{Discussion}\label{sec64}
Our approach may perform much better if the investor can combine
domain knowledge with the statistical modeling that we illustrate
here. We have not done this in the present comparative study because
using a purely empirical analysis of the past returns of these stocks
to build the prediction model~(\ref{eq52}) would be a disservice to the power
and versatility of the proposed approach, which is developed in
Section \ref{sec3} in a general Bayesian framework, allowing the skillful
investor to make use of prior beliefs on the future return vector
${\mathbf r}_{n+1}$ and statistical models for predicting ${\mathbf r}_{n+1}$
from past market data. The prior beliefs can involve both the
investor's and the market's ``views,'' as in the Black--Litterman
approach described in Section \ref{sec22}, for which the market's view is
implied by the equilibrium portfolio. Note that Black and Litterman
model the potential errors of these views by
normal priors whose covariance matrices reflect the
uncertainties. Our Bayesian approach goes one step further to account
for these uncertainties by using the \textit{actual} means and variances
of the portfolio's return in the optimization problem (\ref{eq31}), instead
of the \textit{estimated} means and variances in the plug-in approach.

A portfolio on Markowitz's efficient frontier can be interpreted as
a mini\-mum-variance portfolio achieving a target mean return, or a
maximum-mean portfolio at a given volatility (i.e., standard derivation
of returns). Portfolio managers prefer the former interpretation, as
target returns are appealing to investors. In active portfolio
management [Grinold and Kahn (\citeyear{GK2000})], this has led to the target excess
return
$\widetilde{\mu}_*$ and the optimization problem (\ref{eq61}). The empirical
study in Section \ref{sec61} shows that when the means and covariances of the
stock returns are unknown and are estimated from historical data,
putting these estimates in (\ref{eq61}) may not provide a solution; moreover,
the actual mean of the solution (when it exists) can differ
substantially from~$\widetilde{\mu}_*$.

\section{Concluding remarks}\label{sec7}

The ``Markowitz enigma'' has been attributed to (a)~sampling
variability of the plug-in weights (hence use of resampling to correct for
bias due to nonlinearity of the weights as a function of the mean vector
and covariance matrix of the stocks) or (b) inherent difficulties of
estimation of high-dimensional covariance matrices in the plug-in
approach. Like the plug-in approach, subsequent refinements that
attempt to address (a) or (b) still follow closely
Markowitz's solution for efficient portfolios,
constraining the unknown mean to equal to some target returns. This tends
to result in relatively low information ratios when no or limited short selling
is allowed, as noted in Sections \ref{sec43} and \ref{sec6}. Another difficulty with the
plug-in and shrinkage approaches is that their measure of
``risk'' does not account for the uncertainties in the parameter
estimates. Incorporating these uncertainties via a Bayesian approach
results in a~much harder stochastic optimization problem
than Markowitz's deterministic optimization problem, which we
have been able to solve by introducing an additional parameter $\eta$.

Our solution of this stochastic optimization problem opens up new
possibilities in extending Markowitz's mean--variance portfolio
optimization theory to the case where the means and covariances of
the asset returns for the next investment period are unknown. As pointed
out in Section \ref{sec53}, our solution only requires the posterior mean
and second moment matrix of the return vector for the next period,
and one can combine the Black--Litterman-type expert views with
statistical modeling to
develop Bayesian or empirical Bayes models with good predictive
properties, for example, by using (\ref{eq52}) with suitably chosen ${\mathbf
x}_{i,t-1}$.

\section*{Acknowledgment} We thank the referees for their helpful
comments and suggestions.

\begin{supplement}
\sname{Supplement}
\stitle{Matlab implementation of the NPEB method\\}
\slink[doi,text={10.1214/ 10-AOAS422SUPP}]{10.1214/10-AOAS422SUPP}
\slink[url]{http://lib.stat.cmu.edu/aoas/422/supplement.zip}
\sdatatype{.zip}
\sdescription{The source code of our approach is provided.}
\end{supplement}
%


\printaddresses

\end{document}